\documentclass[manuscript]{aastex}
\shorttitle{\textit{Suzaku} X-ray Observations of the Fermi Bubbles (II)}
\shortauthors{Tahara et al.}

\begin{document}

\title{\textit{Suzaku} X-ray Observations of the Fermi Bubbles: \\
Northernmost Cap and Southeast Claw Discovered with MAXI-SSC\\
}

\author{M. Tahara\altaffilmark{1}, J. Kataoka\altaffilmark{1}, Y. Takeuchi\altaffilmark{1}, T. Totani\altaffilmark{2}, Y. Sofue\altaffilmark{3}, J. S. Hiraga\altaffilmark{4}, H. Tsunemi\altaffilmark{5}, Y. Inoue\altaffilmark{6},  M. Kimura\altaffilmark{6}, C. C. Cheung\altaffilmark{7}, and S. Nakashima\altaffilmark{8}}

\email{taharamasaya@ruri.waseda.jp}

\altaffiltext{1}{Research Institute for Science and Engineering, Waseda University, 3-4-1, Okubo, Shinjuku, Tokyo 169-8555, Japan}
\altaffiltext{2}{Department of Astronomy, The University of Tokyo, Bunkyo-ku, Tokyo 113-0033, Japan}
\altaffiltext{3}{Institute of Astronomy, The University of Tokyo, Mitaka, Tokyo 181-0015, Japan}
\altaffiltext{4}{Research Center for the Early Universe, Graduate School of Science, The University of Tokyo, Bunkyo-ku, Tokyo 113-0033, Japan}
\altaffiltext{5}{Department of Earth and Space Science, Osaka University, Osaka 560-0043, Japan}
\altaffiltext{6}{Institute of Space and Astronautical Science, JAXA, 3-1-1 Yoshinodai, Chuo-ku, Sagamihara, Kanagawa 252-5210, Japan}
\altaffiltext{7}{Space Science Division, Naval Research Laboratory, Washington, DC 20375, USA}
\altaffiltext{8}{Department of Physics, Graduate School of Science, Kyoto University, Kitashirakawa Oiwake-cho, Sakyo-ku, Kyoto 606-8502, Japan}

\begin{abstract}
We report on \textit{Suzaku} observations of large-scale X-ray structures 
possibly related with the Fermi Bubbles obtained in 2013 with a total 
duration of \(\simeq\) 80 ks. The observed regions were the: (i) northern 
cap (N-cap; $l$ \(\sim\) 0$^{\circ}$, 45$^{\circ}$ $<$ $b$ $<$ 55$^{\circ}$) seen 
in the Mid-band (1.7$-$4.0 keV) map recently provided by MAXI-SSC 
and  (ii) southeast claw  (SE-claw; $l$ \(\sim\) 10$^{\circ}$, $-$20$^{\circ}$ $<$ $b$ $<$$-$10$^{\circ}$) 
seen in the ROSAT all-sky map and MAXI-SSC Low-band 
(0.7$-$1.7 keV) map. In each region, we detected 
diffuse X-ray emissions which are represented by a three 
component plasma model consisting of an unabsorbed thermal component 
({\it kT} \(\simeq\) 0.1 keV) from the Local Hot Bubble, absorbed 
{\it kT} = 0.30$\pm$0.05 keV emission representing the Galactic 
Halo (GH), and a power-law component due to the isotropic cosmic X-ray background radiation. 
The emission measure of the GH component in the SE-claw shows an excess by a factor of 
\(\simeq\) 2.5 over the surrounding emission at 2$^{\circ}$ away. 
We also found a broad excess in the 1.7$-$4.0 keV 
count rates across the N-cap after compiling other archival data from \textit{Suzaku} and \textit{Swift}. 
The spectral stacking analysis of the N-cap data 
indicates the presence of 
another thermal component with {\it kT} = 0.70$^{+0.22}_{-0.11}$ keV. The temperature of 
$kT$ $\simeq$ 0.3 keV of the Galactic Halo is higher than the ubiquitous value of 
$kT$ $\simeq$ 0.2 keV near the Fermi Bubbles, and can be even higher ($\sim$ 0.7 keV). 
We discuss our findings in the context of bubble-halo interaction. 

\end{abstract}

\keywords{acceleration of particles --- cosmic rays --- Galaxy: halo --- X-rays: ISM}

\section{INTRODUCTION}

The Large Area Telescope (Atwood et al. 2009) on-board the {\it Fermi} 
satellite discovered the huge \(\gamma\)-ray structures, 
the so-called \textquotedblleft Fermi Bubbles\textquotedblright.
They extend north and south centering on the Galactic Center for about 5 kpc in each direction, 
with a longitudinal width \(\sim\) 40$^\circ$ (Dobler et al. 
2010; Su et al. 2010; Ackermann et al. 2014).
It has been suggested that these bubbles have relatively sharp edges 
(with an angular width of $\simeq$ 3.4$^\circ$$^{+3.7^\circ}_{-2.6^\circ}$) and are symmetric with respect to 
the Galactic plane and the minor axis of the Galactic disk.

Large-scale structures which may be associated with 
the \(\gamma\)-ray Fermi Bubbles are also known in other wavelengths.
The \textquotedblleft WMAP haze\textquotedblright, known as the 
microwave excess and characterized by a flat-spectrum 
is spatially correlated with the \(\gamma\)-ray bubbles at $|$$b$$|$ $<$ $35$$^{\circ}$.
In X-rays, the ROSAT all-sky survey (RASS) provides all-sky
images in the 0.5$-$2.0 keV range (Snowden et al. 1995). 
Possible signatures related with the north bubble can be seen in 
the ROSAT map (e.g., Su et al. 2010, Figs. 18 \& 19 therein), together 
with two sharp edges in the south that seem to trace the Fermi bubble below 
the Galactic disk (e.g., Fig. 1 in Wang 2002; Fig. 6 in 
Bland-Hawthorn \& Cohen 2003; also, Figs. 18 \& 19 in Su et al. 2010).
In the 0.75 keV (R45 band) map, we can see the the North Polar Spur (NPS) 
which is the large-scale soft X-ray enhancement 
corresponding with a part of the radio Loop I structure.
Several authors argued that the NPS and the Loop I may be related to 
the wind activity of the Scorpio-Centaurus OB association at a
distance of 170 pc (Egger \& Aschenbach 1995) or may be a
nearby supernova remnant (Berkhuijsen et al. 1971). In contrast, 
Sofue (2000, and references therein) asserted that the NPS is a 
large-scale outflow from the GC and with a total energy of 
\(\sim\) 10$^{55}$$-$10$^{56}$ erg released on a timescale of 
\(\sim\) 10$^{7}$ yr. 

Kataoka et al. (2013; Paper-I herein) carried out \textit{Suzaku} X-ray observations 
of high-Galactic latitude regions (43$^\circ$ \(< \mid b \mid <\) 54$^\circ$) 
positioned across the north-east and the southern-most edges of the Fermi bubbles 
in an AO7 program in 2012. In all pointings, they found diffuse X-ray 
emission that was well-reproduced by a three-component spectral model 
consisting of unabsorbed thermal emission (temperature $kT \simeq 0.1$\,keV) 
from the Local Hot Bubble (LHB), absorbed $kT \simeq 0.3$\,keV thermal emission related 
to the Galactic Halo (GH), and a power-law component at a level consistent with 
the cosmic X-ray background (CXB). The emission measure (EM) of the 0.3\,keV 
plasma decreases by $\simeq 50\%$ toward the inner regions of the 
north-east bubble without a notable temperature change (in this context, see also 
Fujita et al. 2014). They argued that the presence of a large amount of neutral 
matter absorbing the X-ray emission as well as the existence of the $kT \simeq 0.3$\,keV 
gas can be naturally interpreted as a weak shock driven by the bubbles' expansion 
in the surrounding medium with velocity $v_{\rm exp} \sim 300$\,km\,s$^{-1}$ 
compressing the GH gas to form the NPS feature. Also they suggested that the pressure 
and energy of the non-thermal plasma filling the bubbles are in a rough equilibrium with 
that of the thermal plasma surrounding the bubbles. 

Following Paper-I, we started a new project proposing a sequence of X-ray observations 
and systematic analysis of archival data provided by \textit{Suzaku} X-ray Imaging Spectrometer (XIS; Koyama et al. 2007)
and the \textit{Swift} (Gehrels et al. 2004) X-ray Telescope (XRT; Burrows et al. 2005). Our goal was to find evidence for characteristic X-ray 
structures possibly related with the Fermi bubbles within Galactic longitudes 
$|$$l$$|$ $<$ $30$$^{\circ}$ and latitudes $|$$b$$|$ $<$ $60$$^{\circ}$. Four \textit{Suzaku} observations 
were conducted as part of an AO8 program in 2013 with a total (requested) duration of 
80 ksec. These observations were motivated by the recent finding by
the Monitor of All-Sky X-ray Image (MAXI; Matsuoka et al. 2009) Solid-state Slit Camera (SSC; Tsunemi et al. 2010)
of two prominent X-ray structures in the northernmost and southeast regions of the bubbles, as detailed below. The new MAXI-SSC images and \textit{Suzaku} observations 
are described in section 2. The analysis of diffuse X-ray emission for the AO8 observations data and 
the archival \textit{Suzaku} and \textit{Swift} data is given in section 3. In section 4 we discuss 
a possible scenario accounting for the obtained results.

\section{OBSERVATIONS AND ANALYSIS}

\subsection{MAXI-SSC all sky maps}

The MAXI mission is the first astronomical payload installed on 
the Japanese Experiment Module - Exposed Facility (JEM-EF or Kibo-EF) 
on the International Space Station (ISS). MAXI scans almost the entire (90\%) sky twice 
in every ISS orbit (\(\sim\) 90 minutes).
MAXI carries two detectors -- the Gas Slit Camera (GSC; Mihara et al. 2011) and 
the aforementioned SSC.
Given the spectral resolution of its X-ray CCD, all-sky X-ray images in the energy range of 
0.5$-$12 keV can be generated with the MAXI-SSC with a field of view of 
1.5$^{\circ}$ \(\times\) 90$^{\circ}$. The SSC is one of the 
best instruments to study large diffuse X-ray structures
as recently demonstrated in a study of the
Cygnus superbubble (Kimura et al. 2013). 

While a detailed X-ray
spectral study of the Fermi bubbles using the MAXI-SSC data is 
ongoing, all-sky maps in various X-ray energy bands have 
been prepared for various uses. 
Figure 1 shows the X-ray sky maps obtained with the MAXI-SSC taken from 
2009 August 18 to 2012 February 01 in the Low-band (L-band = 0.7$-$1.7 keV: $top$-$left$) 
and Mid-band (M-band = 1.7$-$4.0 keV: $top$-$right$) (Kimura et al. 2013). 
These maps are exposure-time corrected,
while no background is subtracted. Whereas the global structure seen 
in the L-band image is essentially consistent with the ROSAT 
all-sky maps taken at 0.75 and 1.5 keV (Snowden et al. 1995), 
the M-band image shows prominent structures which are unseen 
in the L-band image. Specifically, we found a ``cap'' or 
clump-like structure appearing at the northernmost part of the Fermi bubble 
at high Galactic latitudes 45$^{\circ}$ $<$ $b$ $<$ 55$^{\circ}$ 
(hereafter, N-cap). Also there appears to be a NPS-like feature visible to the 
southeast in the L-band image. A ``claw'' is seen at low Galactic latitudes 
$-$20$^{\circ}$ $<$ $b$ $<$$-$10$^{\circ}$ (hereafter SE-claw) in the L-band image, 
which is also seen in the  $ROSAT$ 0.75 keV 
map. These new all-sky maps from MAXI-SSC motivated us to 
obtain follow-up observations using \textit{Suzaku} as detailed below. 
\textit{Suzaku} is suitable for investigating such faint diffuse 
X-ray emission because of its low instrumental background and superior 
energy resolution especially below 1 keV.

\subsection{\textit{Suzaku} Observations and Data Reduction}

As a part of an AO8 program in 2013, we carried out 
four \textit{Suzaku} observations of $\simeq$ 20 ksec each, pointed 
``on'' and ``off'' the (i) N-cap and (ii) SE-claw regions (Fig. 1) with a 
total (requested) duration of 80 ksec.  The \textit{Suzaku} observation 
logs and pointing positions are summarized in Table 1. 
The \textit{Suzaku} satellite (Mitsuda et al. 2007) is equipped with four 
X-ray telescopes (XRT; Serlemitsos et al. 2007), each carrying a focal-plane
X-ray CCD camera (X-ray Imaging Spectrometer, XIS; Koyama et al. 2007).
One of the XIS sensors is a back-illuminated (BI) CCD (XIS1), and the other
three XIS sensors are front-illuminated (FI) ones (XIS0, XIS2, and XIS3).
Because the operation of XIS2 ceased in 2006 November due to contamination
by a leaked charge, we use only the other three CCDs.
The XIS was operated in the normal full-frame clocking mode with the
3 \(\times\) 3 or 5 \(\times\) 5 editing mode. Although \textit{Suzaku} 
also carries a hard X-ray detector (Takahashi et al. 2007), 
we do not use the data collected by its PIN and GSO instruments because 
the thermal emission we describe below is too faint to be detected at 10 keV 
and no statistically significant excesses over the CXB were found with these 
PIN/GSO detectors.

The lower panels of Figure 1 show our \textit{Suzaku} XIS fields of view 
(FOVs; \(\sim\) 18' \(\times\) 18') 
onto the MAXI-SSC image for the series of northernmost and southeast 
observations of the bubbles. Green and white circles denote our previous AO7 pointings 
as published in Paper-I and our newly obtained AO8 pointings, respectively, while yellow dashed 
curves indicate the boundary of the Fermi Bubbles as drawn by 
Su et al. (2010). Note that in the N-cap region, there are some regions 
with previous \textit{Suzaku} pointings and the corresponding focal centers 
of these observations are denoted in the figure by cyan circles; the analysis of these 
archival data are discussed below.

We conducted all data reduction and analysis with HEADAS software version 6.14
using the calibration database (CALDB) released on 2013 August 13.
First, we combined the cleaned event data of the two editing modes using XSELECT.
The data corresponding to epochs of low-Earth elevation
angles (less than 20$^\circ$ during both night and day) were then excised, as well periods 
when the \textit{Suzaku} satellite was passing through (and 500 sec after) the 
South Atlantic Anomaly.
Moreover, we also excluded the data obtained when the \textit{Suzaku} satellite was
passing through the low Cut-Off Rigidity (COR) of below 6 GV.
Finally, hot and flickering pixels were removed using SISCLEAN (Day et al. 1998).
With these data selection criteria applied, the resulting
total effective exposures for all the observed pointings are summarized in Table 1.

\subsection{XIS Analysis}

Because the BI CCD (XIS1) has higher instrumental background than those of FI CCDs, we extracted 
the XIS images from only the two FI CCDs (XIS0, XIS3).
We made the XIS images in three photon energy ranges of 0.4$-$10 keV,
0.4$-$2 keV, and 2$-$10 keV. In the image analysis, calibration 
sources located at the corners of the CCD chips were excluded.
The images of Non X-ray Background (NXB) were obtained from the night Earth
data using XISNXBGEN (Tawa et al. 2008).
Since the exposure times for the data toward the N-cap and SE-claw (original data) 
were different from that of the NXB, we corrected for the differences in the exposure 
times using XISEXPMAPGEN (Ishisaki et al. 2007) before subtracting them from the original images.
In addition, we implemented a vignetting correction by simulating flat sky images with XISSIM (Ishisaki et al. 2007).
The XIS0 and XIS3 images were then combined, re-binned by a factor 
of 4 (i.e., 4 pixels were combined to 1), then smoothed by a Gaussian function
with \(\sigma\) = 0.07' (Gaussian kernel radius set as 8 in DS9),
and the resultant images are presented in section 3.

With the combined XIS0+3 images, we first ran the source detection algorithm
in XIMAGE (Giommi et al. 1992) to differentiate compact X-ray features from 
diffuse X-ray emission.
Through this approach, bright compact emitters which are not associated
with catalogued sources were detected (all at approximately \(\geq 3\sigma\)).

We used all the (FI and BI) CCDs, namely, XIS0, 1, 3 for the spectral analysis 
to increase the photon statistics. Regarding the bright compact X-ray features 
described above, we defined the source regions as 2' radius circles centered 
on the emission peaks, and the corresponding background regions were set to 
annuli on the same CCD chips with inside and outside radii of 2' and 4', respectively.
We made redistribution matrix files (RMFs) and auxiliary response files
(ARFs) using XISRMFGEN and XISSIMARFGEN (Ishisaki et al. 2007), respectively.
For the diffuse emission analysis, we set the source region to the whole CCD chip 
which remained after excluding all the compact features detected 
at significance levels above 3\(\sigma\), with typical 2' radius circles which is sufficient to avoid contamination from the compact sources.
Using XISSIMARFGEN and new contamination files (released on 2013 August 13), 
the ARF files were created assuming the uniform extension of the diffuse emission 
within 20' radii orbicular regions (giving the ARF area of 0.35 deg$^{2}$).
As opposed to the point source analysis, we subtracted as background, the NXB data 
obtained from the region in the same CCD chip. 

We note that even though the most recent version of XISSIMARFGEN replicates
well the deterioration of the CCD quantum efficiency below 2 keV,
significant differences in the spectral shape of the measured emission continua
between XIS 0, 1, and 3 were found below \(\simeq\) 0.5 keV due to 
contamination\footnote{https://heasarc.gsfc.nasa.gov/docs/astroe/prop\_tools/suzaku\_td/node10.html}.
Hence we ignored the spectral data below 0.6 keV for XIS 0 and 3, and below 
0.5 keV for XIS 1. We also did not use the high energy data above 
8 keV for all CCDs due to low photon statistics. Although we constructed 
light curves for bright X-ray compact features, none of them suggested 
any significant variability within the short exposures obtained.

\section{RESULTS}

\subsection{X-ray Images}

Figure 2 shows the obtained 0.4$-$10 keV \textit{Suzaku}-XIS images (combined XIS0 and 3) for
the N-cap field (N-cap\_on/off) and the SE-claw field (SE\_on/off).
Note the bright features seen in the images near the corners of the CCD chips are spurious sources 
resulting from instrumental artifacts, which were excluded in the spectral analysis.
Although the pointing positions are only $\simeq$2$^{\circ}$ apart in the SE-claw, we can see that the surface brightness of the inner bubble 
(SE\_on) is much higher than that of the SE\_off, with ON-OFF flux differences 
being as large as \(\sim\) 10$\sigma$. In contrast, such large variations in surface brightness 
are not clearly seen in the N-cap regions. 

In the figures, magenta crosses show the
positions of compact X-ray features detected at \(\geq 3\sigma\) level which are
probably associated with catalogued background sources from the NED and SIMBAD databases.
The other compact features detected here with \textit{Suzaku} at high significances that
miss any identifications in these databases are named 
\textquotedblleft src1$-$5\textquotedblright and indicated as red circles in Figure 2.
Given the limited photon statistics and the relatively large PSF of the XIS,
it is difficult to determine whether the detected compact features are point-like 
or extended.
In this context, it is important to estimate the expected number of background 
active galactic nuclei (AGN) 
likely to be found within each XIS FOV. Following Stawarz et al. (2013), 
we estimate that two or three uncatalogued AGN are present within 
each XIS FOV, corresponding to the number of detected compact X-ray features 
without any optical identification in the present data. In order to investigate the possible associations with 
background AGN, we looked into the available mid-infrared (MIR) maps 
of the areas covered by our \textit{Suzaku} exposures using the NASA Widefield 
Infrared Survey Explorer (WISE; Wright et al. 2010) satellite data.
As a result, we found just one WISE counterpart (indicated as a black cross
in Figure 2) which was thus excluded in the spectral analysis described below.

\subsection{X-ray Spectra}

\subsubsection{Compact X-ray Features}

Using XSPEC, we fitted the spectra of all the compact X-ray features which were 
detected at significance levels above \(\geq 3\sigma\) and have no matches in the 
NED, SIMBAD and WISE databases.
Within the given photon statistics, the X-ray spectra were all well-represented
by single power-law models moderated by Galactic absorption. 
Although the nature of these sources are unclear (as there were no apparent matches in the
NED, SIMBAD and WISE databases), they are likely faint background AGN considering the 
estimate above. Their fitted spectral parameters vary widely
with a range of photon spectral indices \(\Gamma\) from 1.2 to 4.4 and unabsorbed 
2$-$10 keV fluxes (0.2$-$3.5) \(\times\) 10$^{-14}$ erg$^{-1}$ cm$^{-2}$ s$^{-1}$, 
albeit with large uncertainties. In the following, we regard these point-like sources 
as most likely unrelated to the diffuse GH structure, and excluded them 
in the diffuse X-ray emission analysis. \\

\subsubsection{Diffuse Emission}

We regarded the whole CCD chip as a source region after excluding all
the compact features mentioned above, without regard to whether or not 
they were previously catalogued X-ray sources.
We can fit the diffuse spectra of all the \textit{Suzaku} data 
with the same three component model, 
APEC1 + WABS *(APEC2 + PL), as done in Paper-I 
(APEC: an emission from collisionally-ionized diffuse gas, WABS: a photo-electric absorption, 
PL; a single power law). In other words, the model consists of an unabsorbed thermal 
component APEC1 corresponding to the Local Hot Bubble 
(LHB) emission and/or contamination from the Solar-Wind Charge Exchange 
(SWCX; Fujimoto et al. 2007), an absorbed thermal component APEC2 corresponding to 
the GH, and an absorbed single power-law component PL due to the isotropic CXB radiation.
With regard to the absorbed component, we fixed the neutral hydrogen column density 
at the Galactic value $N_{H,Gal}$ corresponding to the direction
of each pointing (see Table 2), while we fixed the photon index for the CXB component
at $\Gamma_{\rm CXB}$ = 1.41 (Kushino et al. 2002).
We also carried out the spectral fitting where $N_{H}$
was left free and derived their ratios relative to the Galactic values, $N_{H}$ /$N_{H,Gal}$ (Table 2).
One can see that the values in the southeast are either consistent with 
or even larger than unity within the statistical errors. 
However, in the N-cap pointings (i.e., N-cap\_on/off) the relatively low photon 
statistics only provided upper limits. 
Following Paper-I and the other literature (Willingale et al. 2003; Miller et al. 2008; 
Yao \& Wang 2005; Yao et al. 2009, 2010), we fixed the temperature and abundance of 
the LHB plasma at {\it kT} = 0.1 keV and {\it Z} = {\it Z}$_{\odot}$, respectively. 
The existence of the {\it kT} \(\simeq\) 0.1 keV component 
related to LHB/SWCX is verified by the past \textit{Suzaku} and XMM observations 
of 12 and 26 pointings, respectively 
(Yoshino et al. 2009; Henley et al. 2010; Henley \& Shelton 2013).
The results of the spectral fitting are shown in Figure 3 and summarized in Table 2.

To interpret the OVII (574 eV) emission especially below 0.7 keV 
the {\it kT} \(\simeq\) 0.1 keV component (APEC1) is still important, while the measured 
continua are dominated by the absorbed thermal component (APEC2) below 2 keV.
We investigated the metallicity of this component by fitting the stacked spectra of the
N-cap and SE-claw regions and found that their metallicities are 
{\it Z} = 0.075 $^{+0.03}_{-0.02}$ {\it Z}$_{\odot}$ and 0.717 $^{+2.29}_{-0.33}$ {\it Z}$_{\odot}$, 
respectively. It appears that the metallicity in the SE-claw is higher than that in 
the N-cap, and even higher than that reported in Paper-I (average of \(\simeq\) 0.2 {\it Z}$_{\odot}$ 
at 42$^{\circ}$ $<$ $b$ $<$ 48$^{\circ}$). This may suggest an actual decrease in the metallicity at increasing
Galactic latitudes, but this is inconclusive with the data in hand due to the relatively poor photon 
statistics. We fixed the metallicity of each area at the corresponding value.
Meanwhile, the X-ray spectra above 2 keV are dominated by the power-law component.
Assuming the isotropy of this emission, the PL intensity at 2$-$10 keV is approximately
consistent with the absolute intensity of the CXB, namely
(5.85 \(\pm\) 0.38) \(\times\) 10$^{-8}$ erg cm$^{-2}$ s$^{-1}$ sr$^{-1}$, as determined by Kushino et al. (2002) 
as an average of 91 observation fields, assuming a power-law model with a photon index $\Gamma_{\rm CXB} = 1.41$.
As shown in Table 2, at least at the level well exceeding CXB intensity fluctuations, 
an excess non-thermal emission component related to the Fermi Bubbles is not seen. 
Although the derived intensities seem rather lower than the absolute value determined by 
Kushino et al. (2002), this 
is naturally expected from the large-scale fluctuation of the CXB itself
as concluded in Kushino et al. (2002) and also demonstrated by Yoshino et al. (2009). 

As seen in Table 2, the GH component (APEC2) temperatures of N-cap\_on/off and SE\_on/off 
are almost identical ({\it kT} = 0.30 $\pm$ 0.05 keV). Comparing the fitted parameters of 
SE\_on to those of SE\_off, the EM of APEC2 increases by a factor of \(\geq\) 2, 
while a similar increase was not seen in the EM across the N-cap. Although the EM 
in the N-cap appears rather to decrease between the N-cap\_off to N-cap\_on regions, it can be naturally expected that 
the EM becomes smaller as Galactic latitude increases because the EM describes the amount of material along 
the line-of-sight. In order to check the plasma 
temperature of {\it kT} $\simeq$ 0.3 keV near the N-cap, we investigated the spectra of five other 
\textit{Suzaku} observation pointings in the N-cap area available in the archive (see Figure 1; N-cap1-5), 
with the same analysis methods as described above and these results are shown in Figure 4 \& 5 and 
summarized in Table 2. The absorbed plasma temperatures in the directions of these pointings were also 
{\it kT} = 0.30 $\pm$ 0.05 keV (see details in Section 4).

\subsubsection{Swift-XRT Analysis}

In addition to the analysis of the \textit{Suzaku} archival data as described above, 
we also analyzed 19 archival \textit{Swift}-XRT observations close to the N-cap area 
(see Figure 6; Swift1-19, numbered in order of decreasing Galactic latitude) to confirm 
the existence of the isothermal diffuse $kT$ $\simeq$ 0.3 keV plasma. 
Although the \textit{Swift} exposures were typically $\le$ 5 
ksec and hence much shorter than the \textit{Suzaku} ones, the number of pointings at different 
Galactic coordinates is well-suited to investigate the plasma temperature over many sightlines.
From the XRT archival data, we extracted the spectra of the diffuse X-ray emission 
that remained after excluding all the compact features detected at significance levels above 
3\(\sigma\) with the same analysis method described above for the \textit{Suzaku} data. Using the 0.4$-$5.0 keV data 
in order to avoid the instrumental contaminations, we were able to fit the diffuse spectra of all 
the \textit{Swift} pointings with the same three component APEC1 + WABS *(APEC2 + PL) model
and all the GH component temperatures found were consistent with {\it kT} $\simeq$ 0.3 keV (results 
summarized in Appendix). These results are further considered
in section 4.

\section{DISCUSSION AND CONCLUSIONS}

We reported the results of \textit{Suzaku} X-ray observations
of two prominent structures likely related to the Fermi Bubbles.
The N-cap was recently discovered with MAXI-SSC and only seen in the M-band 
image, whilst the SE-claw is seen both in the ROSAT map and MAXI-SSC L-band 
image. In section 3.2.2, we showed that the diffuse X-ray emission which remained 
after excluding all compact X-ray features detected at \(\geq 3\sigma\) level, 
was well-represented by a three-component plasma model consisting of unabsorbed thermal emission 
from the Local Hot Bubble, absorbed thermal emission related to the Galactic Halo,
and a \(\Gamma \simeq\) 1.4 power-law component expected from the Cosmic X-ray Background.

The temperature of the GH component, $kT$ $\simeq$ 0.3 keV, seen in both N-cap\_on/off and 
SE\_on/off is consistent with that reported in Paper-I. 
The EM of the GH component changes substantially, dropping by about 
a factor of 2.5 from the SE\_on to SE\_off regions which are only 2$^{\circ}$ apart in the SE-claw. 
A similar variation was also seen in the northeast area data presented in Paper~I. In contrast, such an increase in the EM 
of the $kT$ $\simeq$ 0.3 keV plasma is not seen in the N-cap. 

In Figures 7 \& 8, we plot 1.7$-$4.0 keV count rates and temperature {\it kT} 
of the GH component in the N-cap against Galactic latitude (see Appendix for details). 
Figure 7 includes the data for our new AO8 pointings (N-cap\_on/off), 
the previous AO7 pointings (Paper I; N1-8), and also the archival \textit{Suzaku} 
pointings (N-cap1-5). Figure 8 includes the data for the archival \textit{Swift} 
observation pointings (Swift1-19). 
One can see that the temperature of the GH component is uniform at $kT$ $\simeq$ 0.30 \(\pm\) 0.05 keV, 
consistent with the finding presented in Paper-I. Meanwhile, the 1.7$-$4.0 keV count rates (Fig.~8) 
across the N-cap show a broad excess of about 20 \% over its edge, consistent with what is seen in the 
MAXI M-band image (see Fig.~1). Such excess emission in the M-band cannot be 
accounted for by EM variations of the $kT$ $\simeq$ 0.3 keV plasma modeled by an APEC2 
because such low-temperature thermal plasma do not significantly affect count rates above 1.7 
keV. This may suggest that the excess emission 
is due to the presence of an \emph{additional} plasma component, having much higher temperature. 

Assuming the additional higher temperature plasma component, we generated a spectrum by stacking 
three regions showing excess counts in 1.7$-$4.0 keV. As shown in Fig.~9, the stacked spectrum can 
be well-fit by adding an additional plasma component APEC3 with a temperature of 
$kT$ = 0.70 $^{+0.22}_{-0.11}$ keV. By adding this component the $\chi^2$ has been improved from 168.36 (145 d.o.f.) 
to 155.73 (143 d.o.f.) with statistical significance more than 99$\%$ confidence level (see Table 4). 
The ``hotter'' plasma may be formed via a termination shock. 
In fact, similarly weak shocks were generally found 
in some radio galaxies residing in cluster and group environments, such as in NGC~3801 
(Croston et al. 2007), B2~0838+32A (Jetha et al. 2008), HCG~62 (Gitti et al. 2010), 
NGC~5813 (Randall et al. 2011), 4C~+29.30 (Siemiginowska et al. 2012), 3C~452 
(Shelton et al. 2011), and PKS~B1358-113 (Stawarz et al. 2014).

Together with our finding in Paper-I, it appears that there is a ubiquitous diffuse 
$kT$ $\simeq$ 0.3 keV thermal plasma (modeled by APEC2) within a central region of 
Galactic longitude $|$$l$$|$ $<$ $30$$^{\circ}$ and latitude $|$$b$$|$ $<$ $60$$^{\circ}$. 
Such an isothermal plasma may be virialized and only weakly heated from the ubiquitous 
0.2 keV value to 0.3 keV within the Galactic halo and hence irrelevant to the Fermi bubbles. 
However, the positional coincidence of prominent structures, like the NPS, the N-cap,  
and SE-claw seems to support the idea that they are all related to the formation of the
Fermi Bubbles. We believe similar hot plasma and enhanced structures like NPS and N-cap
could be observed also in the south, but too faint to be detected in the
current instruments onboard MAXI/Suzaku. In fact, one see a counter-part of
NPS denoted as South Polar Spur (SPS), as described in detail in Sofue et
al. (2000). The fact that these south structures are considerably fainter
than those in the north may suggest (1) asymmetric outflow due to different
environmental condition and/or (2)  the axis of the Fermi bubbles may be
leaning from perpendicular to the Galactic disk (Kataoka et al. in prep).

Assuming that the 0.3 keV plasma is heated by a shock driven by the bubbles' expansion 
in the surrounding halo, the corresponding velocity is $v_{\rm exp}$ $\sim$ 300 
km s$^{-1}$ and shock Mach number $\mathcal{M} \simeq 1.5$ (Paper I). 
This is consistent with the recent finding of a non-thermal velocity in the X-ray 
absorption line toward 3C~273 whose sightline passes through the neighborhood of 
the Fermi Bubbles (Fang \& Jiang 2014). Such a low expansion velocity is actually 
expected in some theoretical models (e.g., Crocker et al. 2014, Fujita et al. 2014, 
Mou et al. (2014), but see also, Guo \& Mathews 2012, Guo et al. 2012, Lacki 2014). 
Specifically, using hydrodynamical simulations, Mou et al. (2014) 
argued that the plasma temperature around the Fermi Bubbles 
is $kT$ $\simeq$ 0.3 keV in the model that the bubbles were inflated by 
winds launched from the past hot accretion flow in Sgr A$^{\ast}$. 
Moreover, Fig. 1 in Mou et al. (2014) implies that the temperature of thermal 
plasma can be a bit higher at the outermost regions of the bubbles (i.e., at high Galactic 
latitudes) than at lower Galactic latitudes, possibly reflecting the positional 
 difference of expanding velocity near the bow-shock region. This is qualitatively 
similar to what we observed in the  ``N-cap'' in this paper. 
Further extensive studies using the new MAXI-SSC data and more extensive use of archival X-ray observations 
are now ongoing and will be presented elsewhere.

\acknowledgments

We are grateful to the referee for the useful comments helped the improvement of the paper. 
Work by C. C. C. at NRL is supported in part by NASA DPR S-15633-Y.

\clearpage

\appendix

\section{SWIFT ANALYSIS RESULTS}

In this Appendix, we summarize the \textit{Swift} observation logs and pointing positions (Table 5), 
the results of the spectral fitting (Table 6), and the 1.7-4.0 keV count rates of the diffuse emission (Table 7). 
Note that due to the low photon statistics of the \textit{Swift}-XRT data, the metallicities of the APEC2 component were all fixed at 
{\it Z} = 0.2 {\it Z}$_{\odot}$ as in Paper-I.

\clearpage

\begin{figure}
  \begin{tabular}{cc}
    \begin{minipage}{0.5\hsize}
      \includegraphics[width=\linewidth]{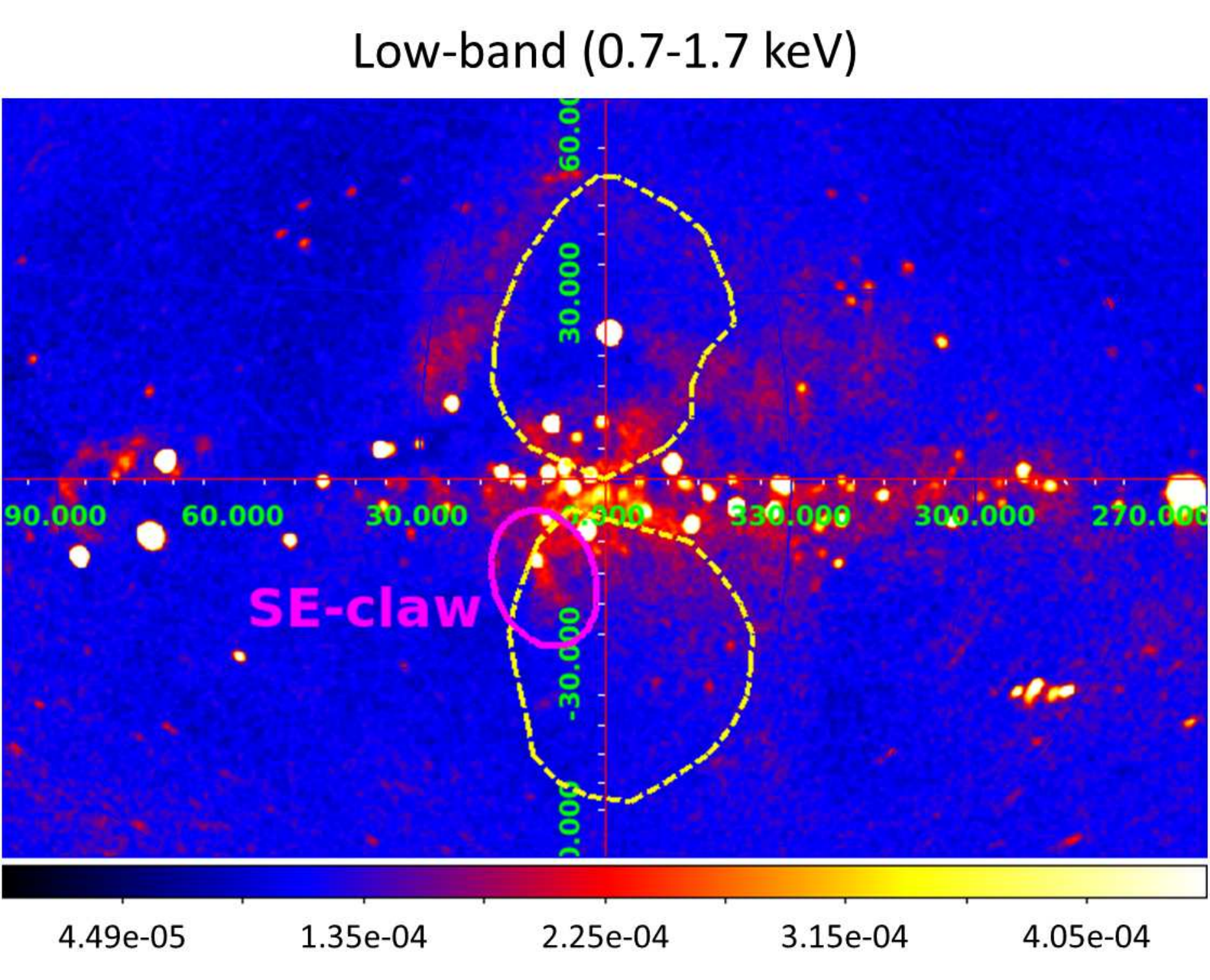}
    \end{minipage}
    \begin{minipage}{0.5\hsize}
      \includegraphics[width=\linewidth]{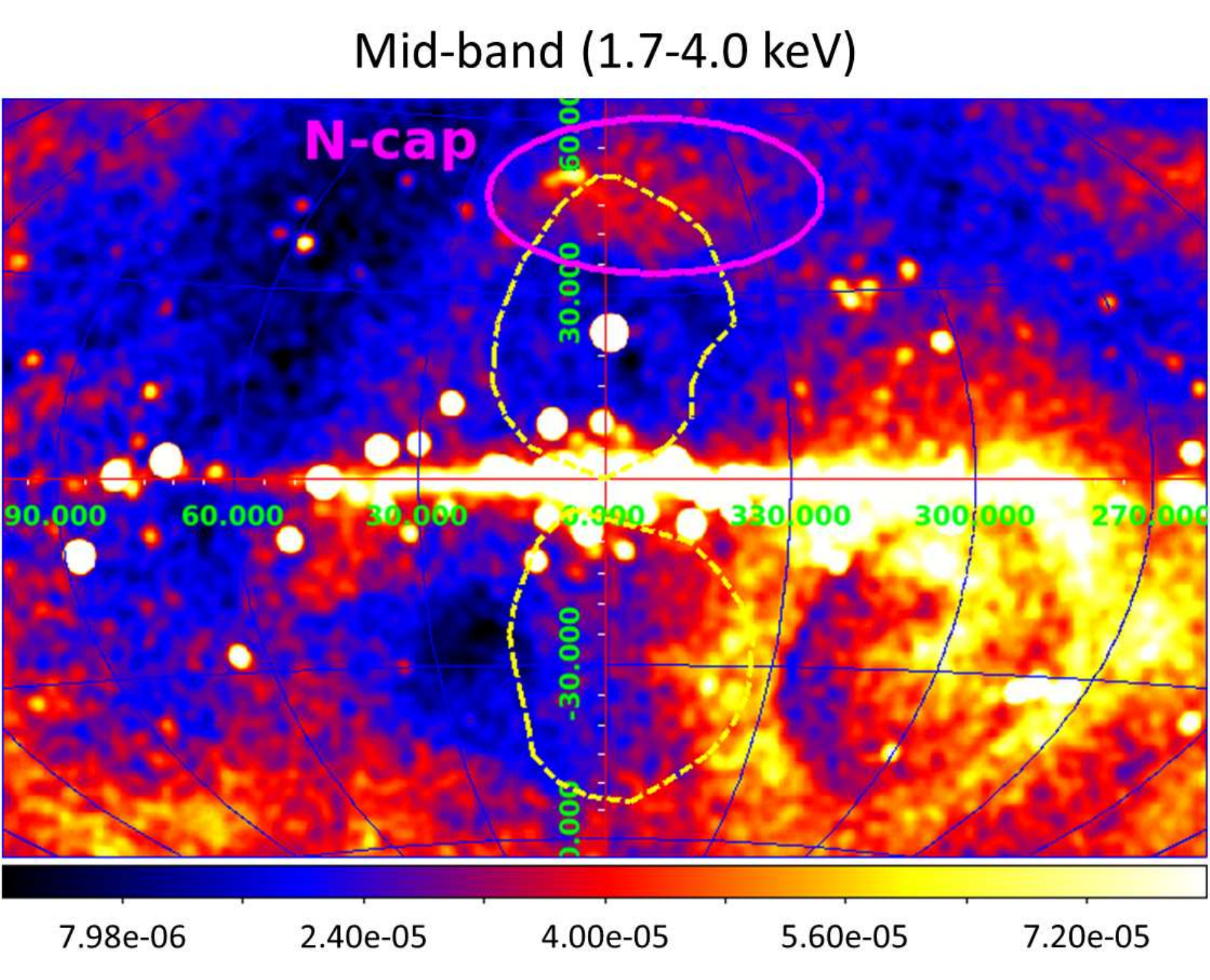}
    \end{minipage}\\
    \begin{minipage}{0.5\hsize}
      \includegraphics[width=\linewidth]{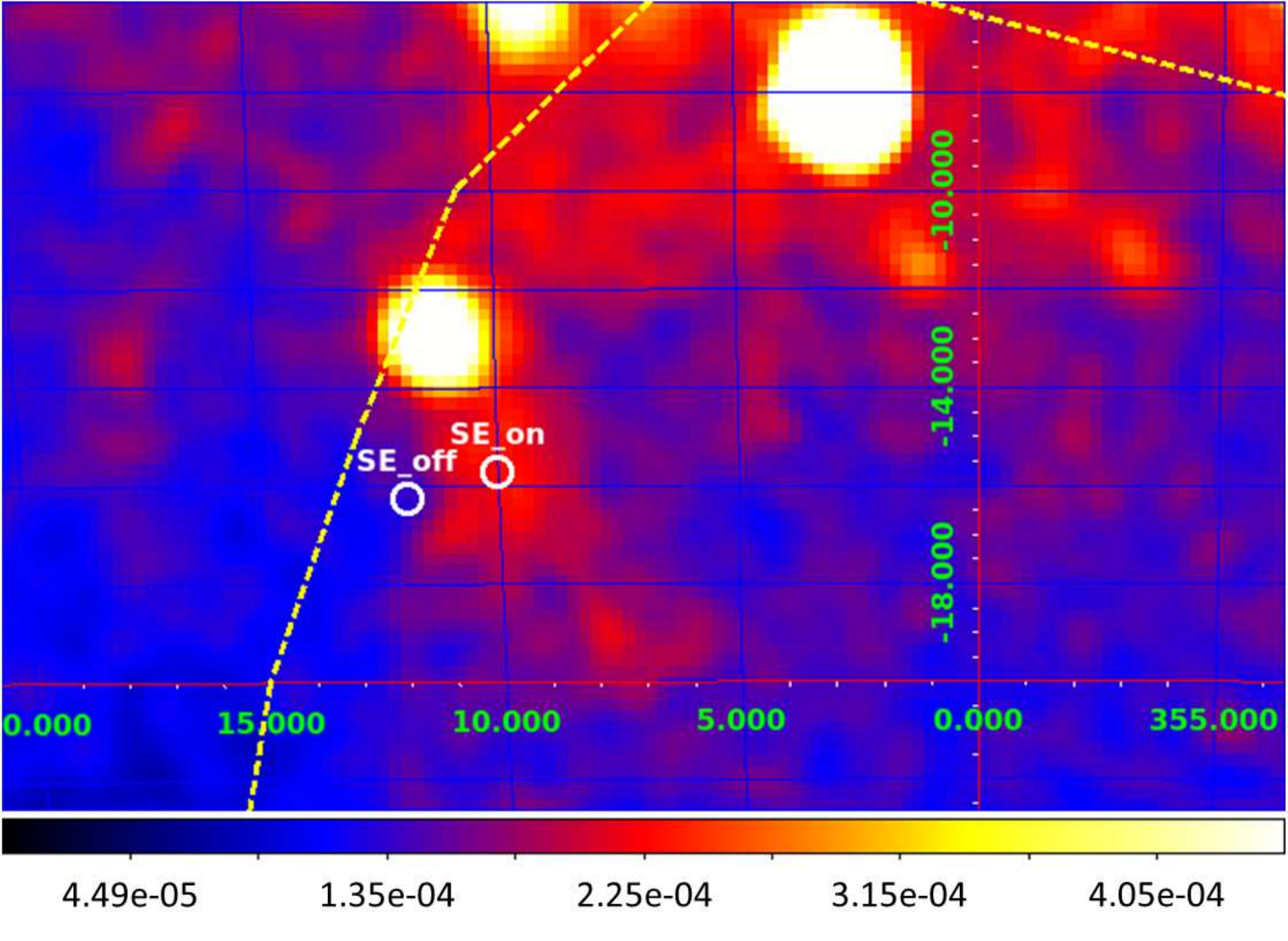}
    \end{minipage}
    \begin{minipage}{0.5\hsize}
      \includegraphics[width=\linewidth]{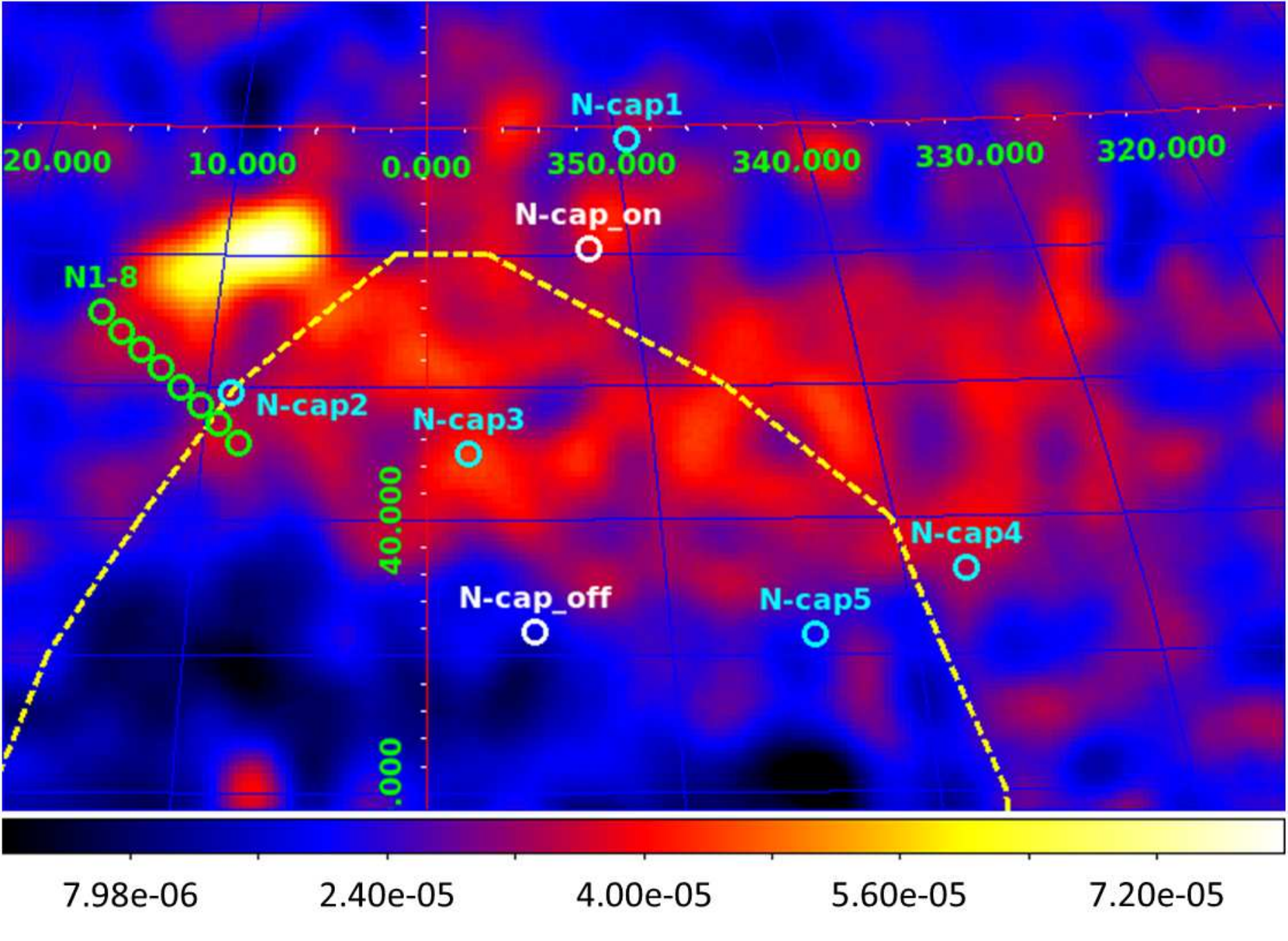}
    \end{minipage}
  \end{tabular}
  \caption{MAXI-SSC images of the Fermi Bubbles in the L-band (0.7$-$1.7 keV: $left$ panels) and in the 
M-band (1.7$-$4.0 keV: $right$ panels). Magenta ellipses denote the two prominent structures, 
N-cap \& SE-claw analyzed in this paper. Yellow dashed lines indicate the boundary of the bubbles as suggested 
in Su et al. (2010). Close-ups of the SE-claw and N-cap are 
shown in {\it Bottom-left} and {\it Bottom-right}, respectively, overlaid 
with the FOVs of the \textit{Suzaku} XIS in AO-8 ($white$ circles), in AO-7 ($green$ circles; Paper-I), and other archival observations ($cyan$ circles). 
The names given to each \textit{Suzaku} FOV are indicated in the figure. All figures are shown in Galactic coordinates. The scale ranges (units of cts s$^{-1}$ cm$^{-2}$) are indicated at the bottom of each panel.
\label{fig1}}
\end{figure}

\clearpage

\begin{figure}
  \begin{center}
    \scalebox{0.7}{
      \begin{tabular}{cc}
        \begin{minipage}{0.5\hsize}
          \includegraphics[width=\linewidth]{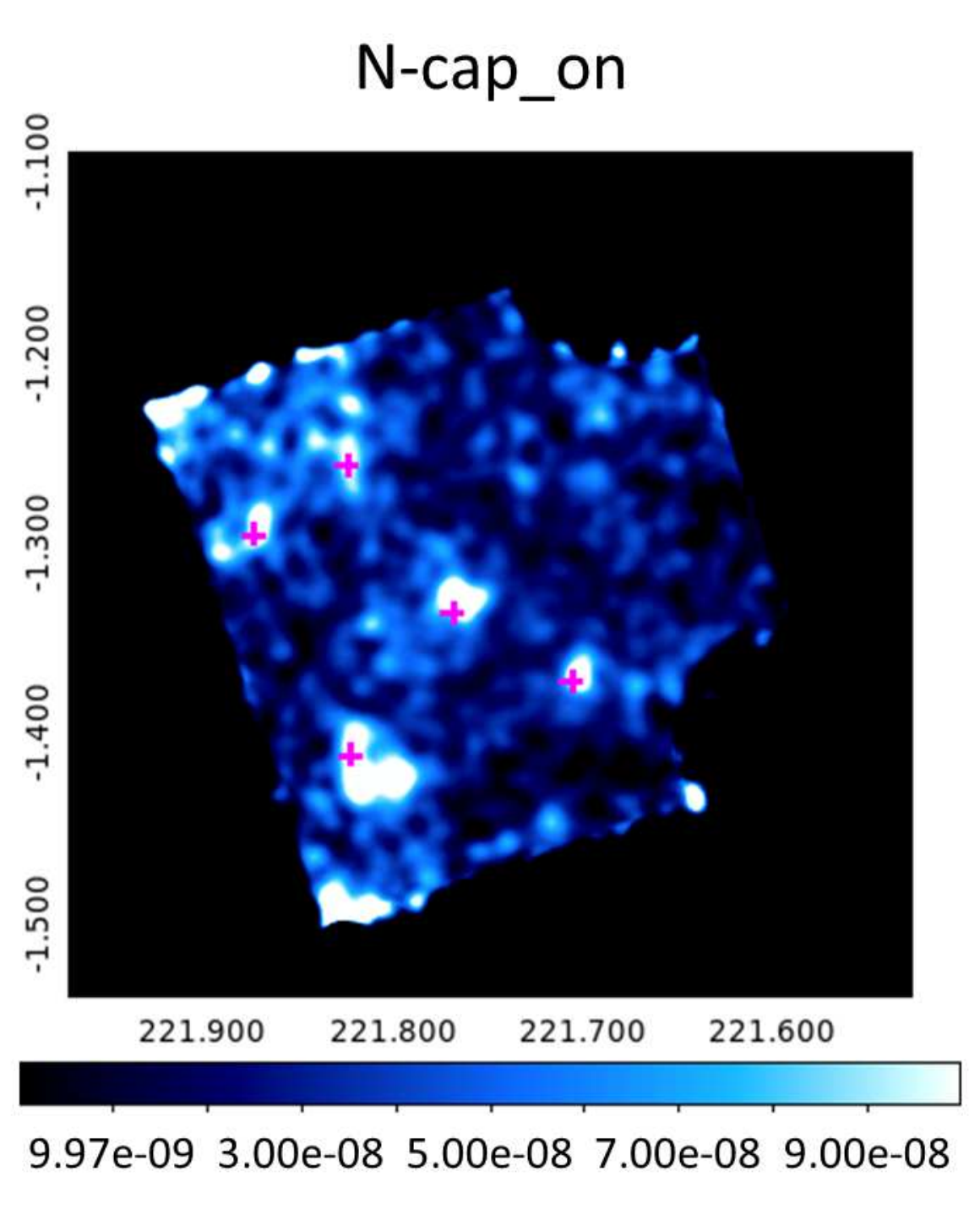}
        \end{minipage}
        \begin{minipage}{0.5\hsize}
          \includegraphics[width=\linewidth]{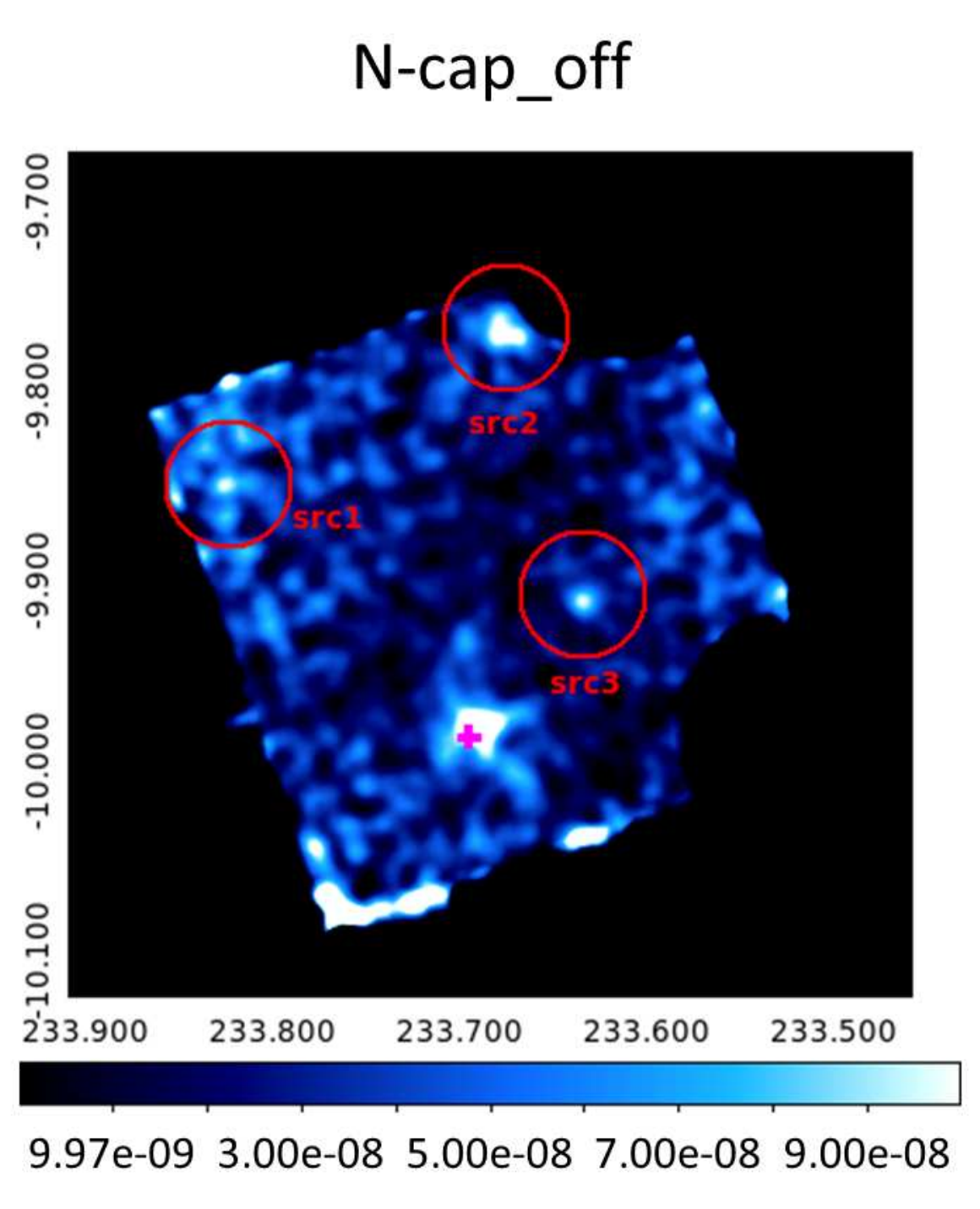}
        \end{minipage}\\
        \begin{minipage}{0.5\hsize}
          \includegraphics[width=\linewidth]{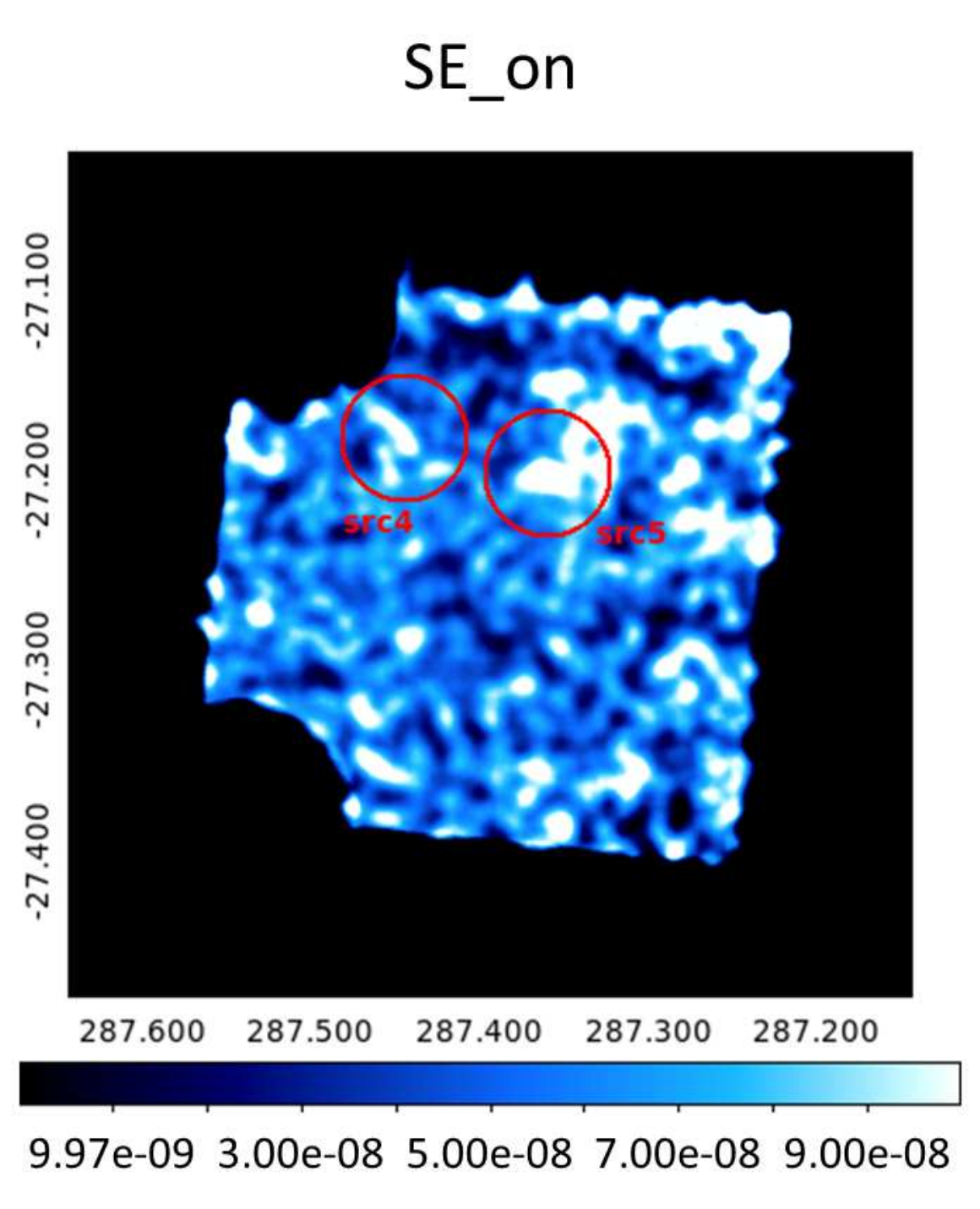}
        \end{minipage}
        \begin{minipage}{0.5\hsize}
          \includegraphics[width=\linewidth]{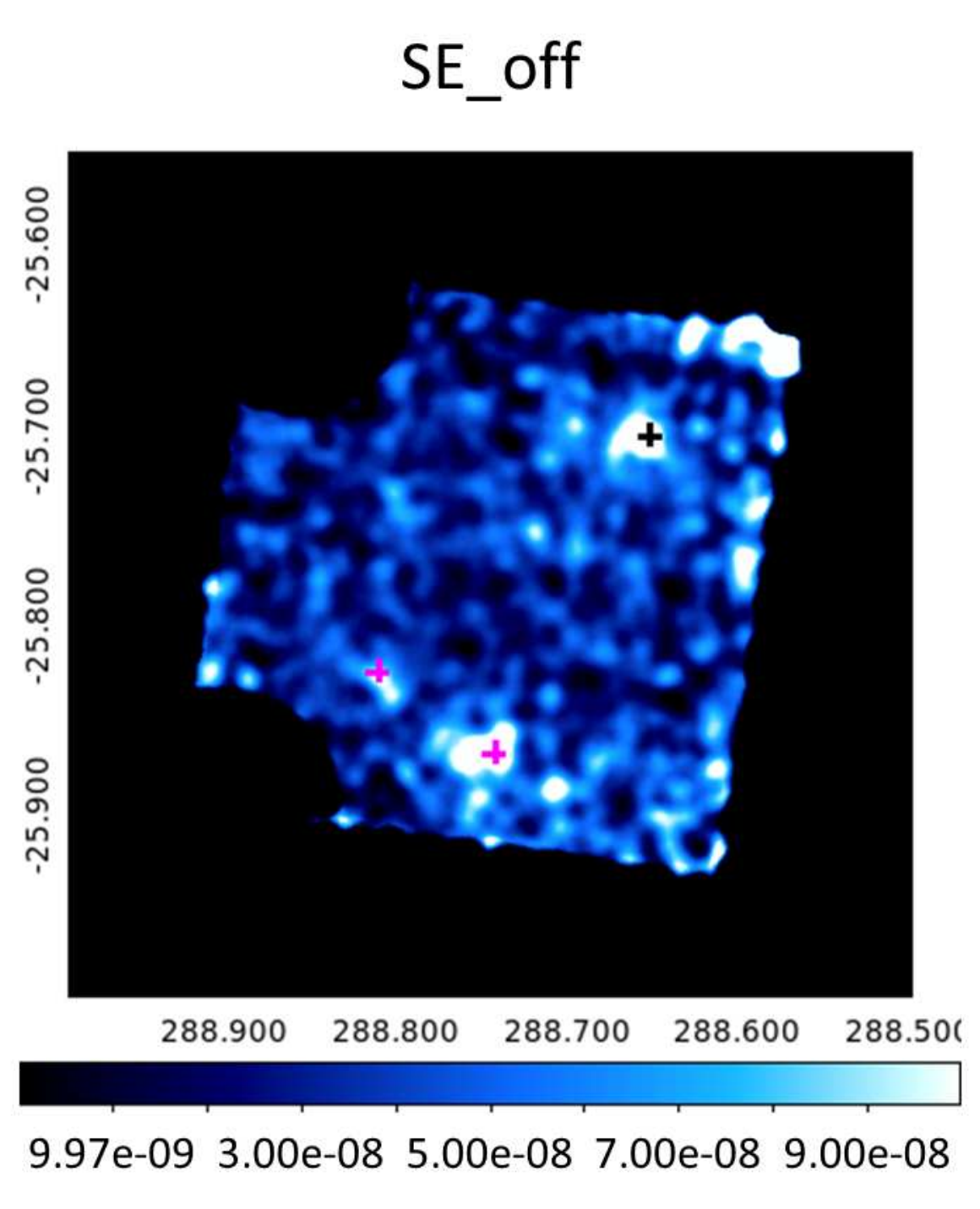}
        \end{minipage}
      \end{tabular}
    }
  \end{center}
    \caption{The 0.4$-$10 keV XIS (XIS 0+3) images of the \textit{Suzaku} pointings of AO-8 (N-cap\_on/off and SE\_on/off indicated in Fig. 1), after vignetting correction and subtraction of the NXB. Uncatalogued X-ray features detected above \(\simeq 3 \sigma\) level are denoted as src1-5; the corresponding source extraction regions are indicated by red circles. The size of the source extraction regions were all \(2'\), chosen to avoid contamination from nearby sources. The bright X-ray features that are catalogued sources are marked with magenta crosses. In addition, the bright X-ray feature correlating with the source seen in WISE data is marked with a black cross. All figures are shown in Equatorial coordinates (J2000) and with the same color scale (units of cts pixel$^{-1}$ where 1 pixel = 2.08'' \(\times\) 2.08'').
    {\it Top-left}: N-cap\_on.
    {\it Top-right}: N-cap\_off.
    {\it Bottom-left}: SE\_on.
    {\it Bottom-right}: SE\_off.\label{fig2}}
\end{figure}


\begin{figure}
  \begin{tabular}{cc}
    \begin{minipage}{0.5\hsize}
      \includegraphics[width=\linewidth]{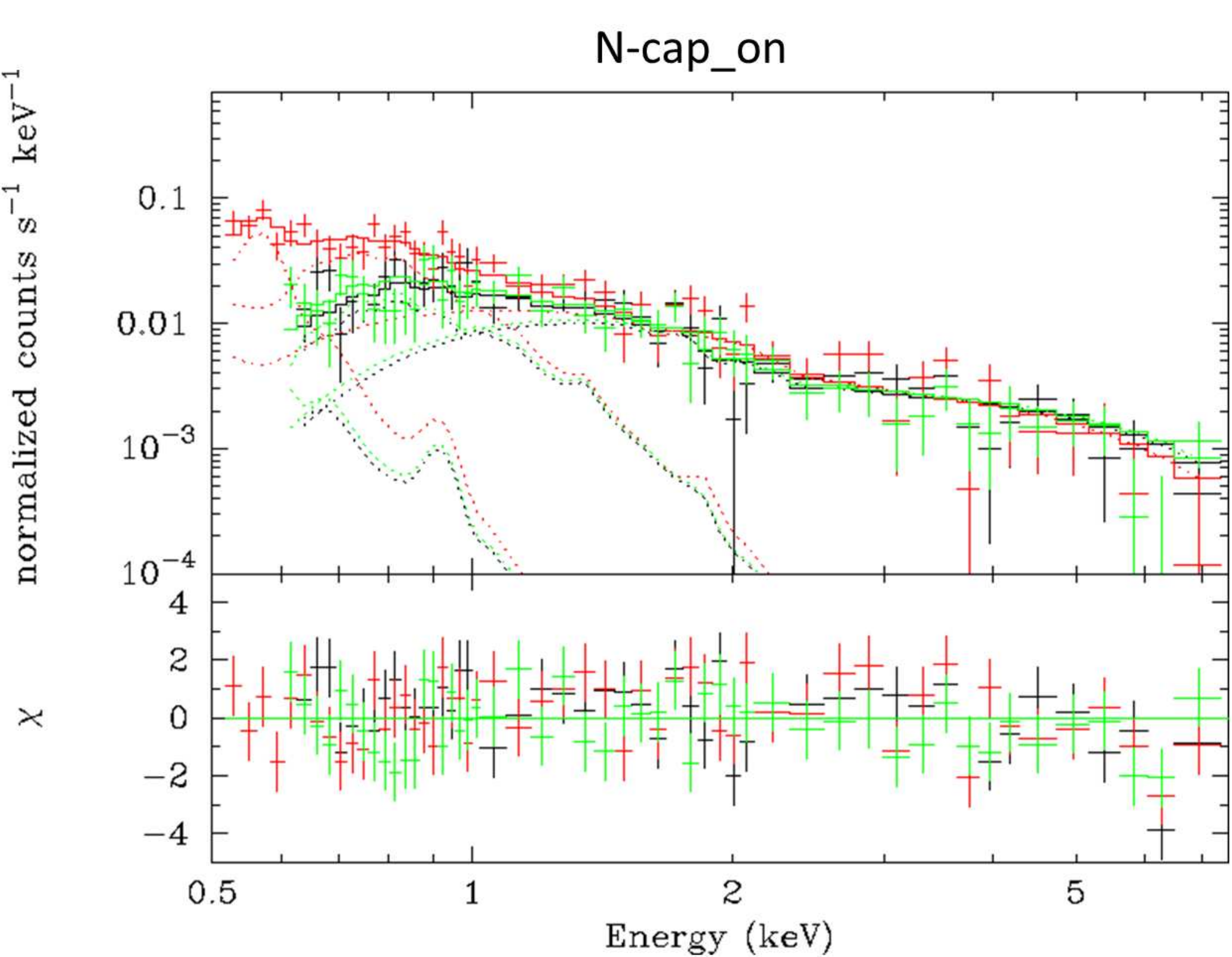}
    \end{minipage}
    \begin{minipage}{0.5\hsize}
      \includegraphics[width=\linewidth]{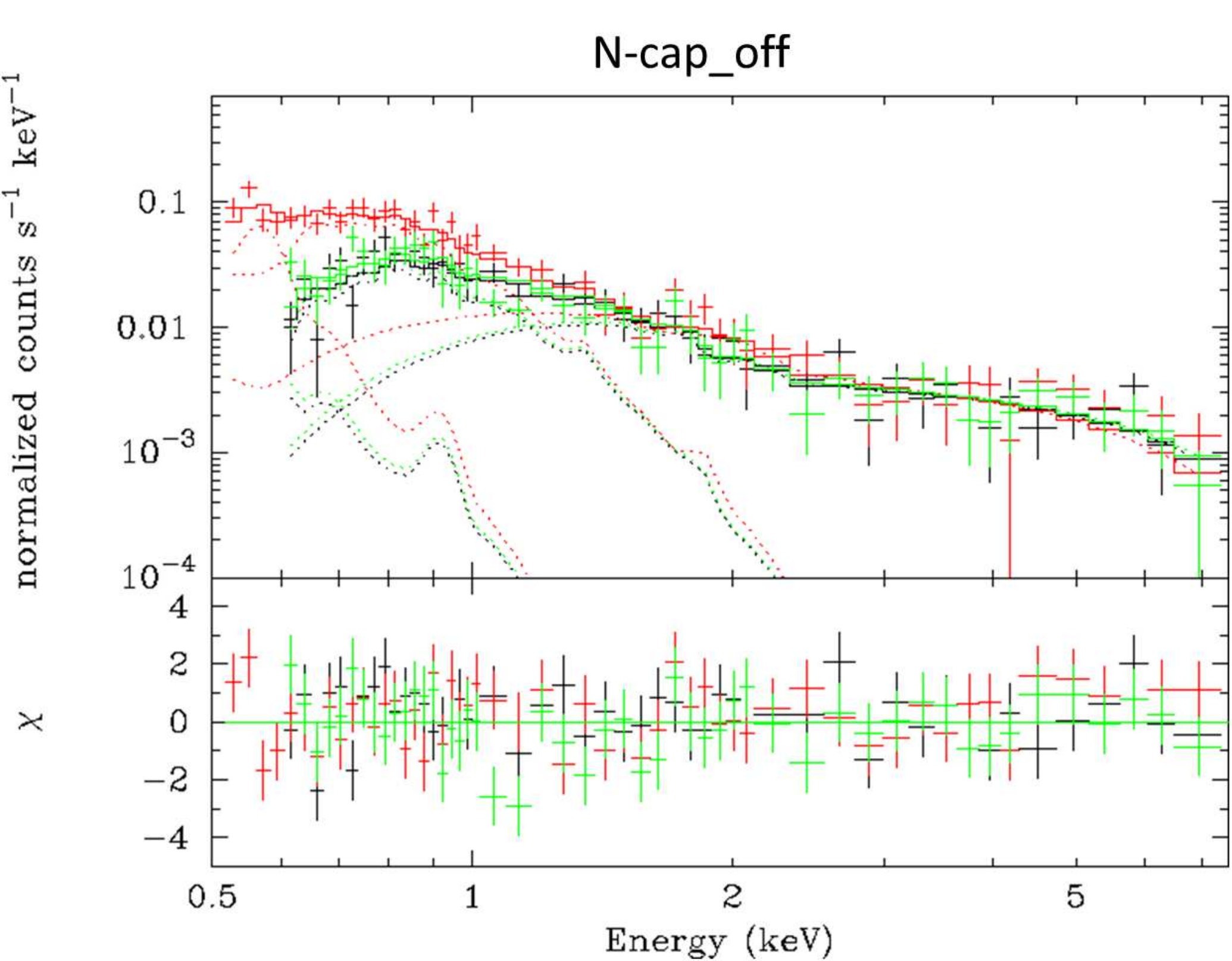}
    \end{minipage}\\
    \begin{minipage}{0.5\hsize}
      \includegraphics[width=\linewidth]{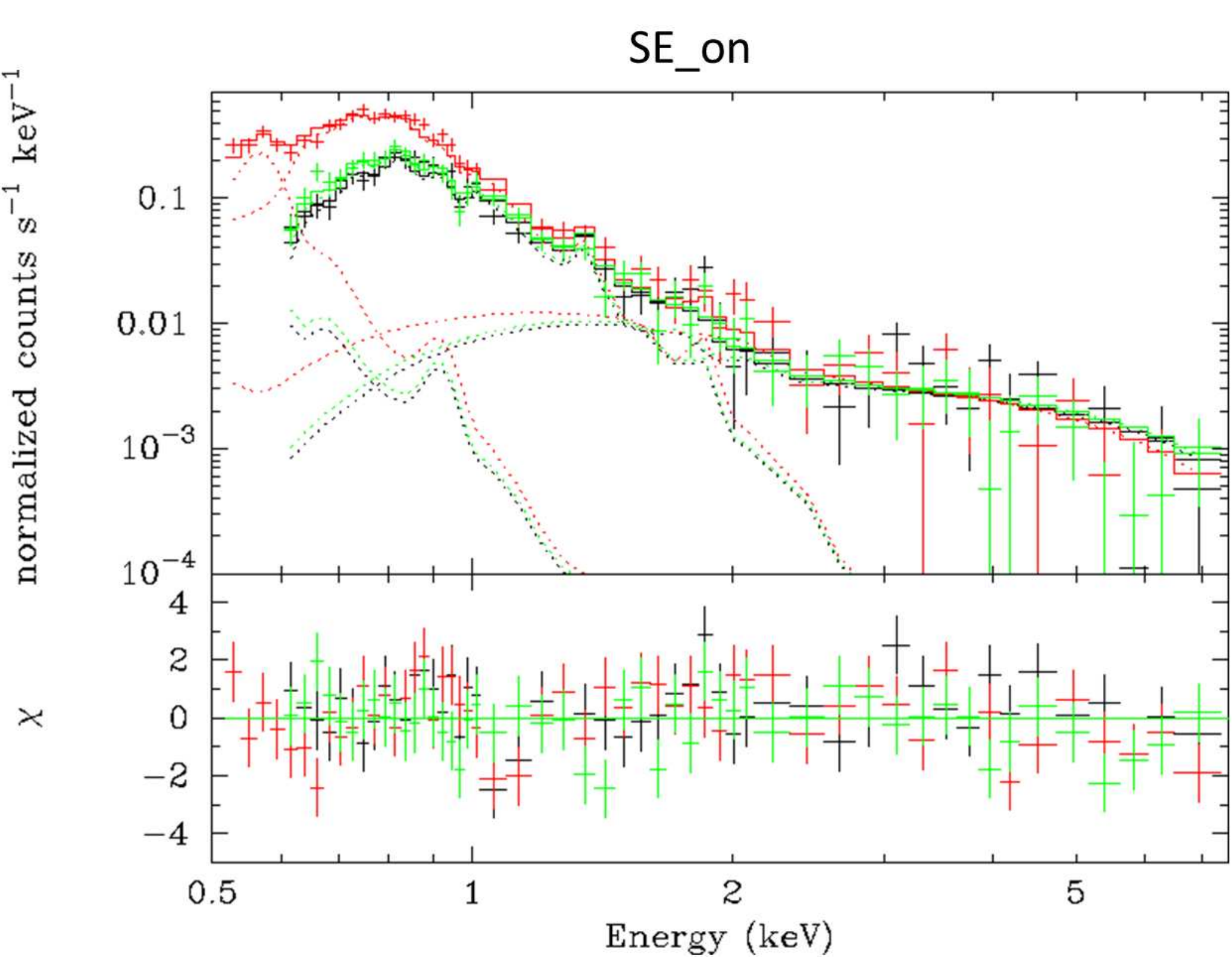}
    \end{minipage}
    \begin{minipage}{0.5\hsize}
      \includegraphics[width=\linewidth]{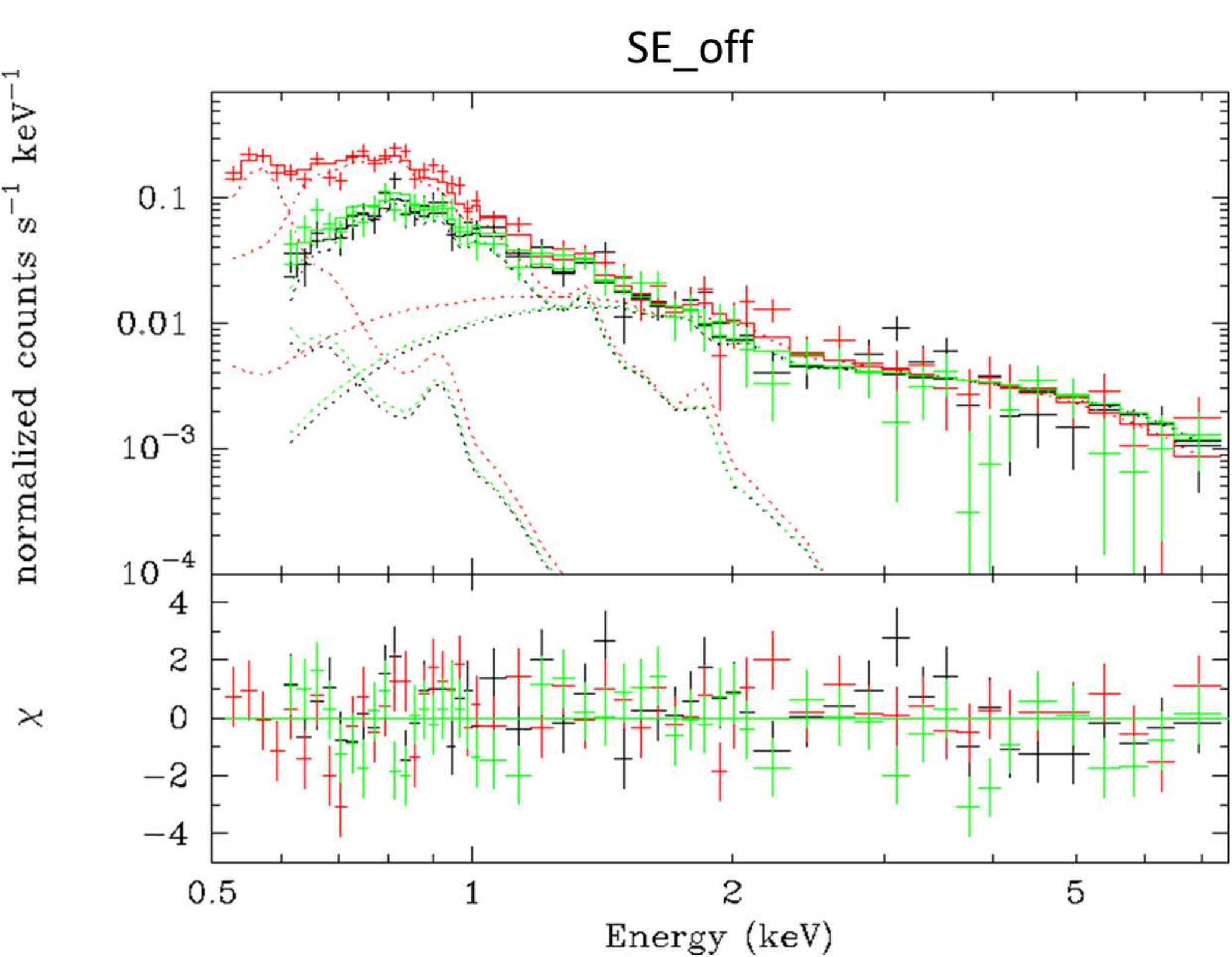}
    \end{minipage}
  \end{tabular}
  \caption{The XIS spectra of the diffuse emission component for the northernmost and southeast \textit{Suzaku} pointings, together with the best-fit model curves (APEC1 + WABS *(APEC2+ PL)) and residuals. The LHB component ({\it kT} \(\simeq\) 0.1 keV; APEC1) dominates the lowest 0.5$-$0.6 keV range, thermal emission related to the GH ({\it kT} \(\simeq\) 0.3 keV; APEC2) dominates the 0.6$-$1.5 keV range, and a power-law component from the CXB dominates above 1.5 keV (\(\Gamma \simeq\) 1.4; PL). XIS 0 data/fits are shown in black, XIS 1 in red, and XIS 3 in green.
    {\it Top-left}: N-cap\_on.
    {\it Top-right}: N-cap\_off.
    {\it Bottom-left}: SE\_on.
    {\it Bottom-right}: SE\_off.\label{fig3}}
\end{figure}

\begin{figure}
  \begin{center}
    \scalebox{0.7}{
      \begin{tabular}{cc}
        \begin{minipage}{0.5\hsize}
          \includegraphics[width=\linewidth]{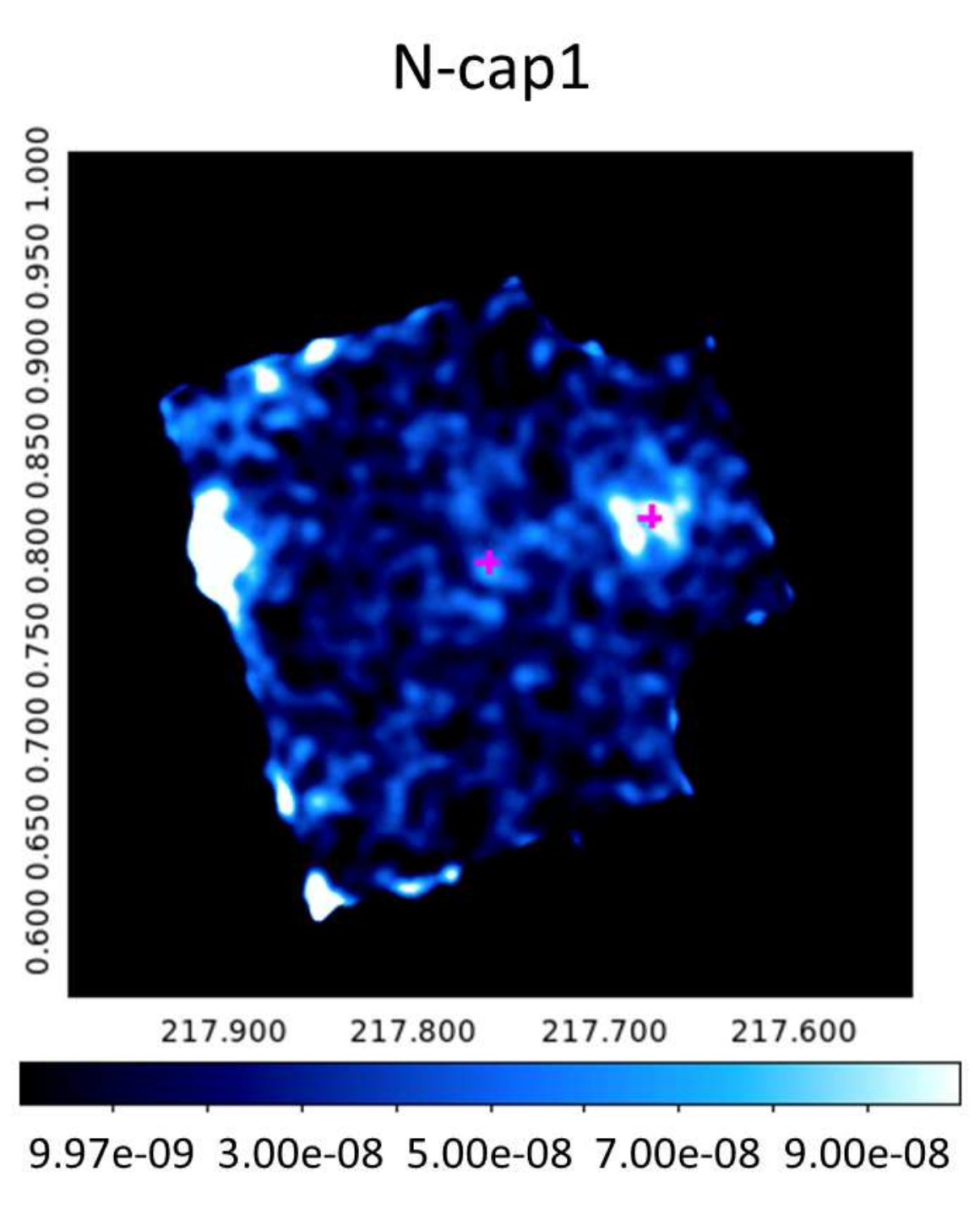}
        \end{minipage}
        \begin{minipage}{0.5\hsize}
          \includegraphics[width=\linewidth]{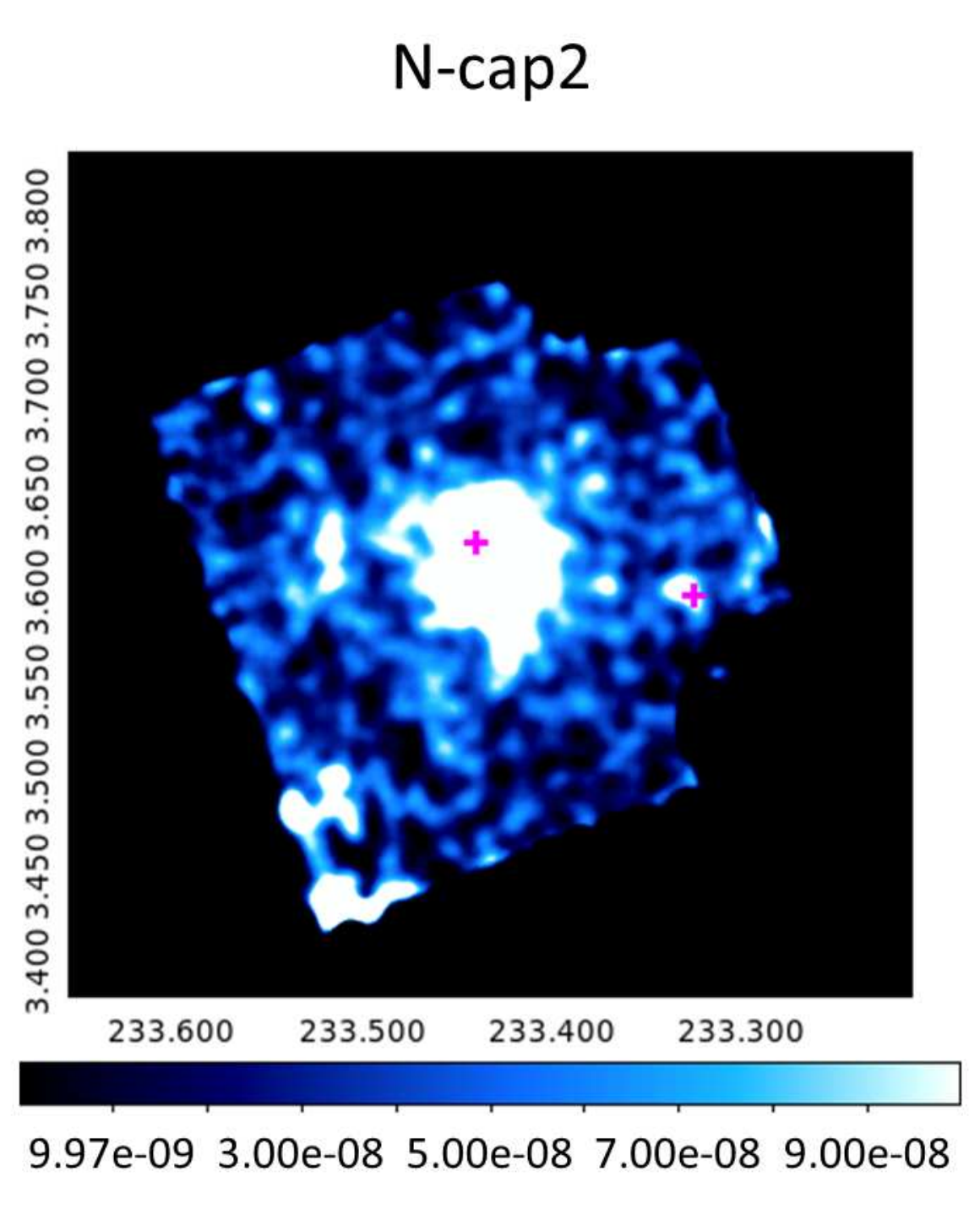}
        \end{minipage}
        \begin{minipage}{0.5\hsize}
          \includegraphics[width=\linewidth]{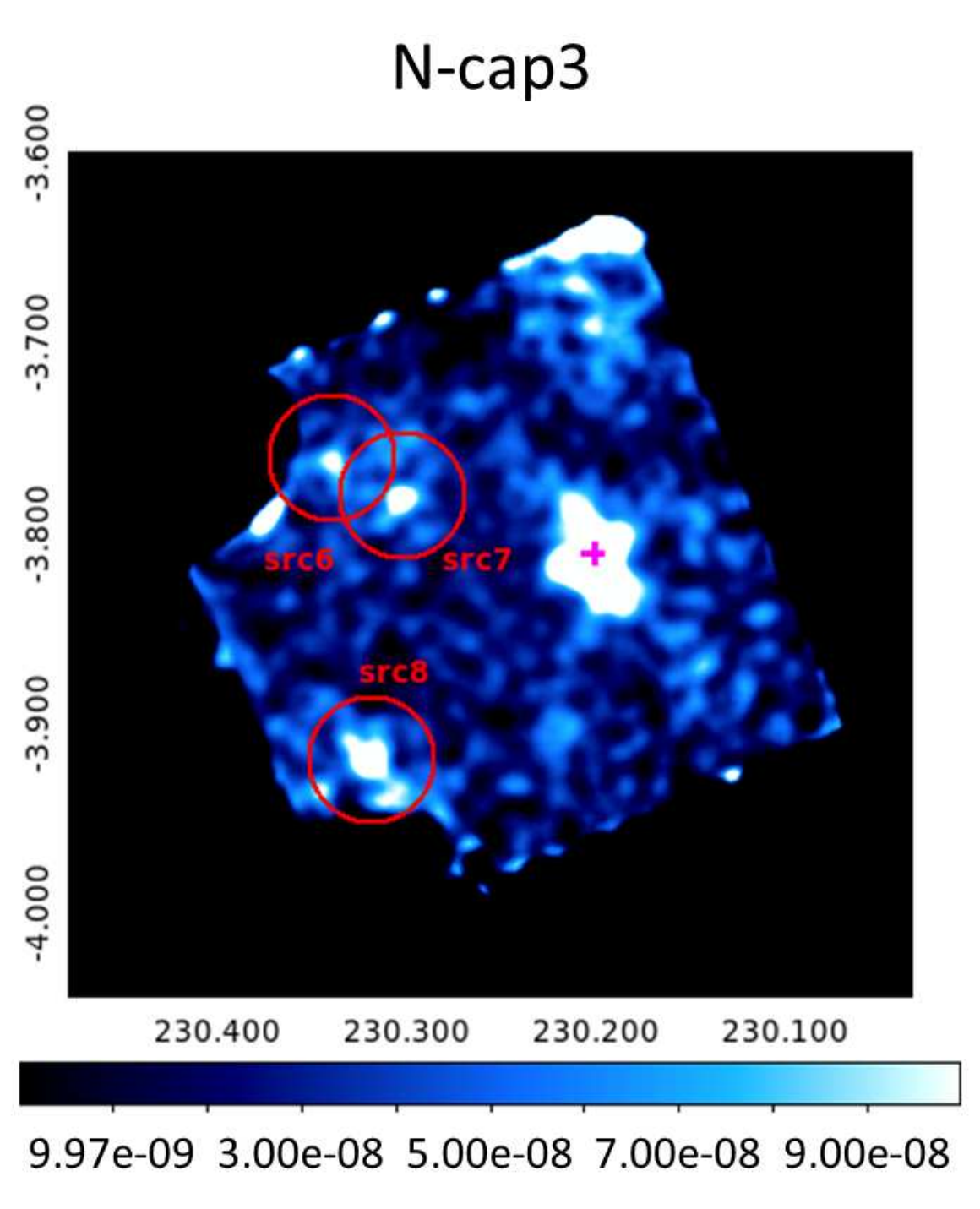}
        \end{minipage}\\
        \begin{minipage}{0.5\hsize}
          \includegraphics[width=\linewidth]{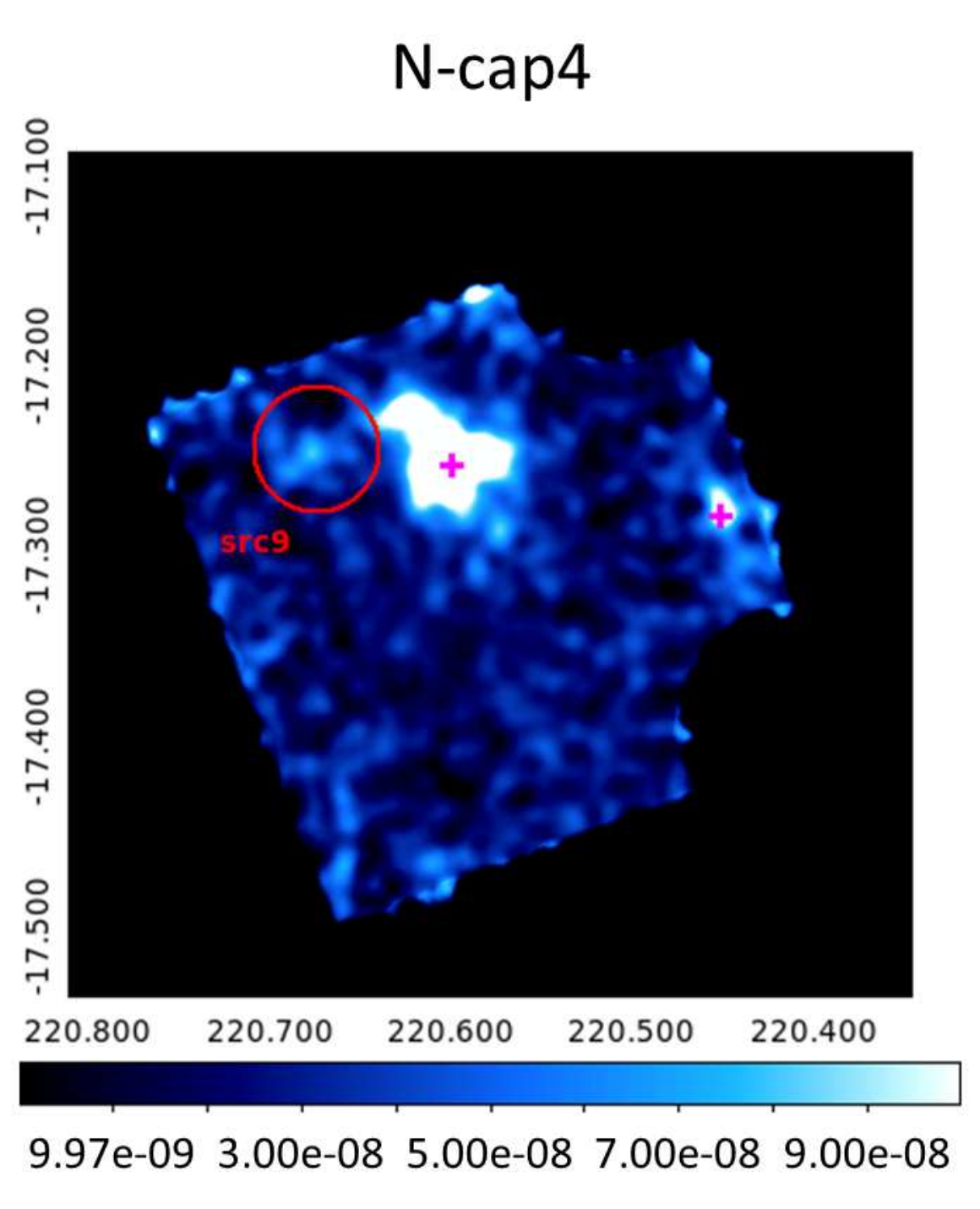}
        \end{minipage}
        \begin{minipage}{0.5\hsize}
      \includegraphics[width=\linewidth]{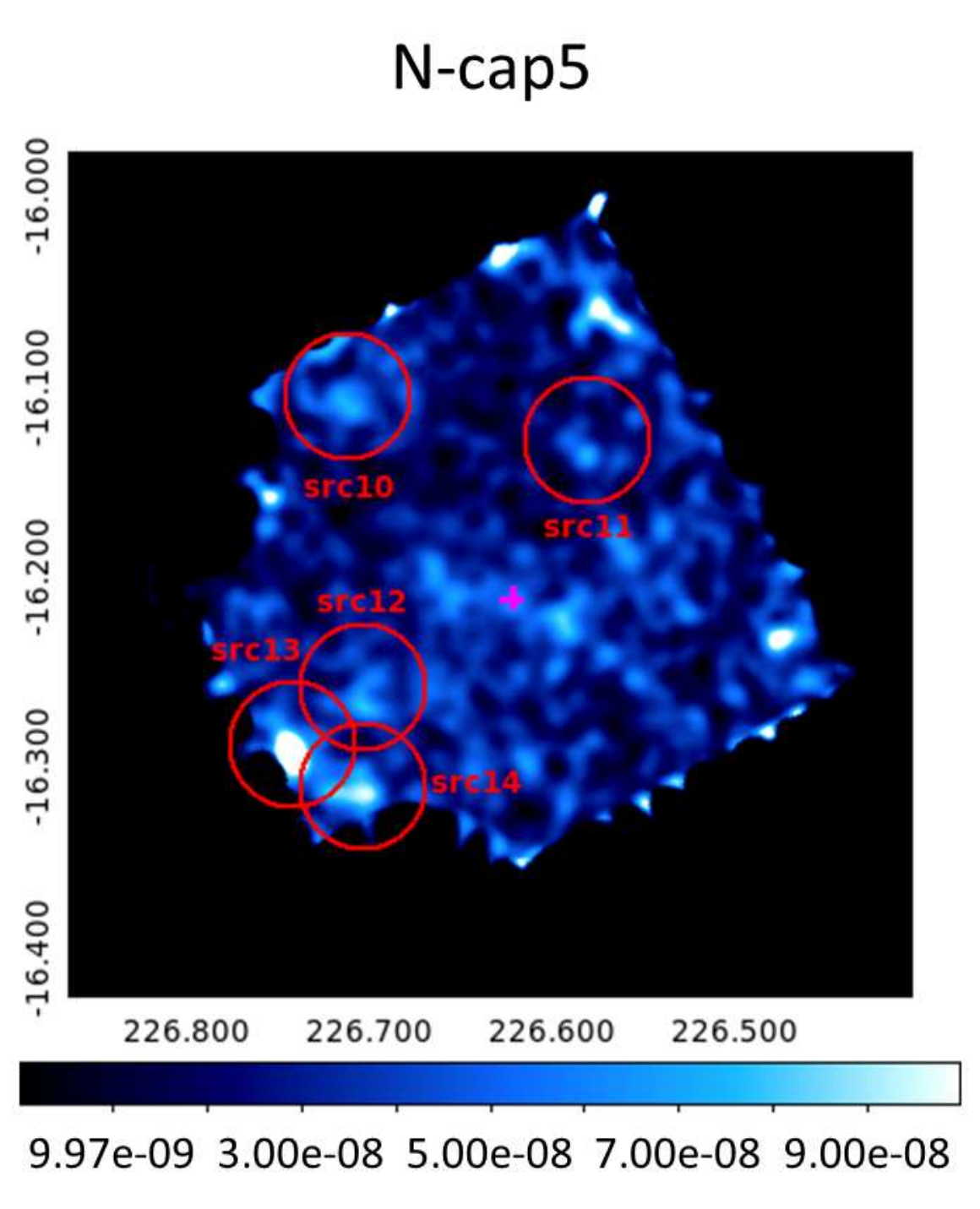}
        \end{minipage}
      \end{tabular}
    }
  \end{center}
  \caption{As in figure 2, but for the archival \textit{Suzaku} pointings in the northernmost structure. Uncatalogued X-ray features detected above \(\simeq 3 \sigma\) level are denoted as src6-14; the corresponding source extraction regions are indicated by red circles.
    {\it Top-left}: N-cap1.
    {\it Top-right}: N-cap2.
    {\it Middle-left}: N-cap3.
    {\it Middle-right}: N-cap4.
    {\it Bottom}: N-cap5.\label{fig4}}
\end{figure}

\begin{figure}
  \begin{tabular}{ccc}
    \begin{minipage}{0.5\hsize}
      \includegraphics[width=\linewidth]{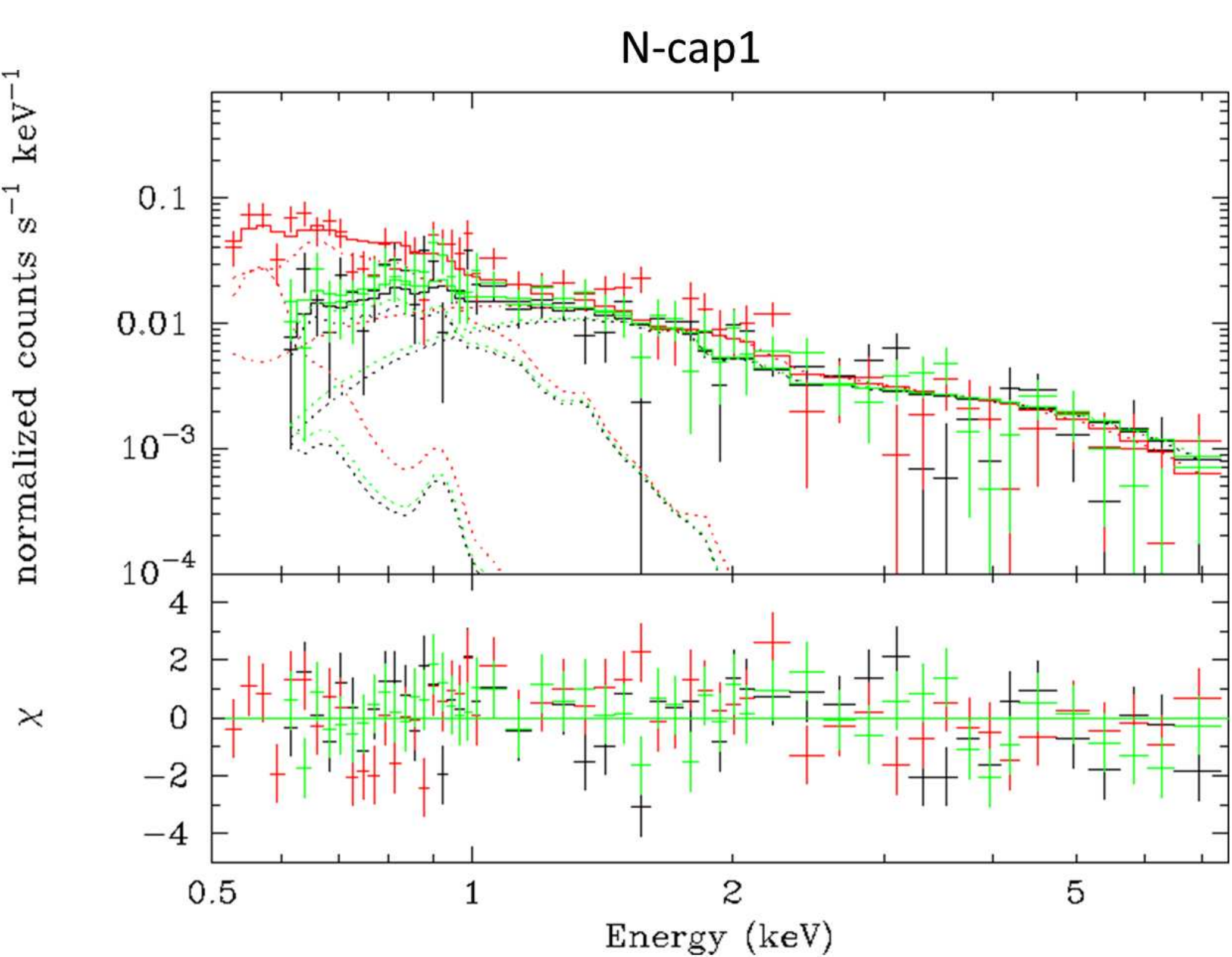}
    \end{minipage}
    \begin{minipage}{0.5\hsize}
      \includegraphics[width=\linewidth]{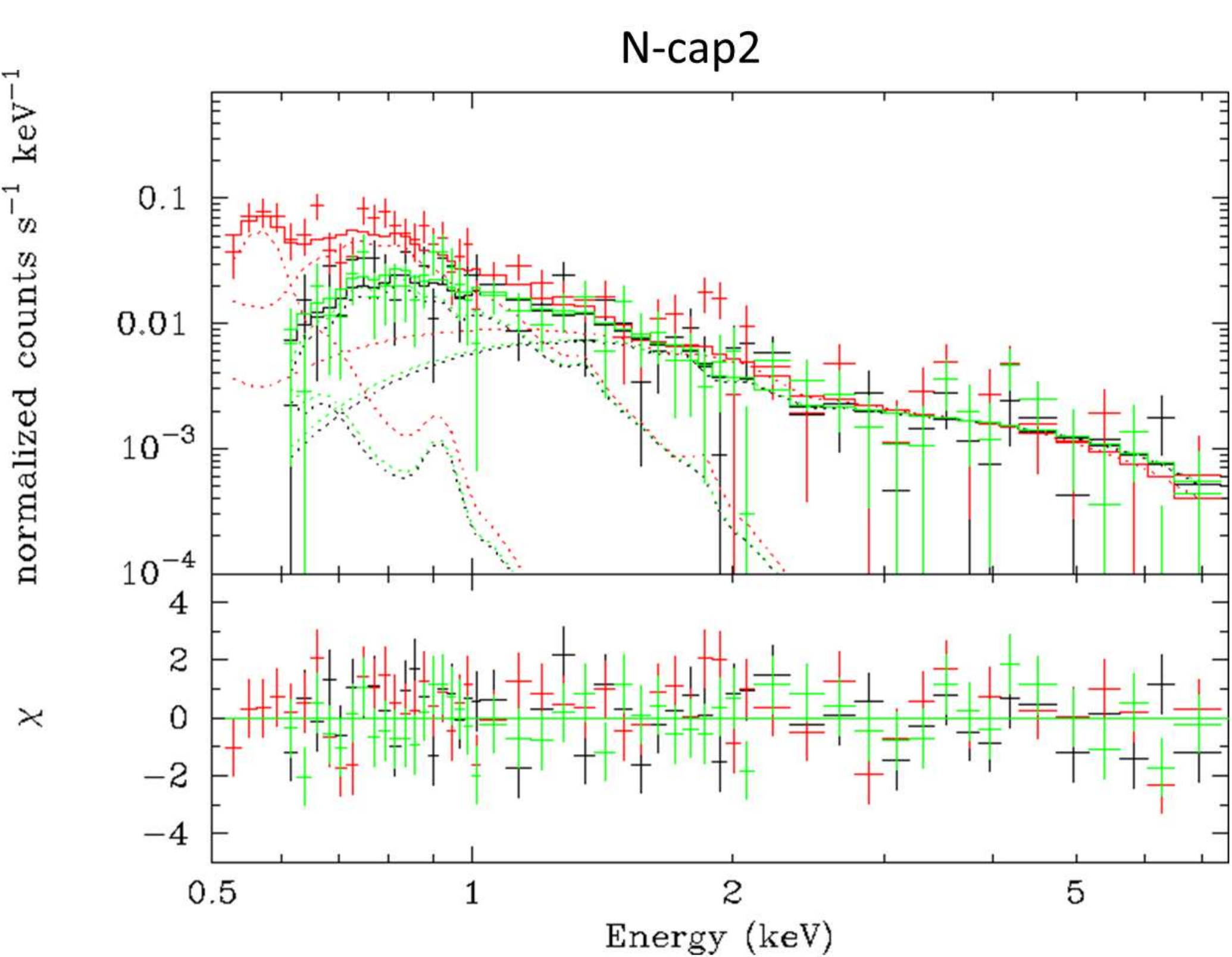}
    \end{minipage}\\
    \begin{minipage}{0.5\hsize}
      \includegraphics[width=\linewidth]{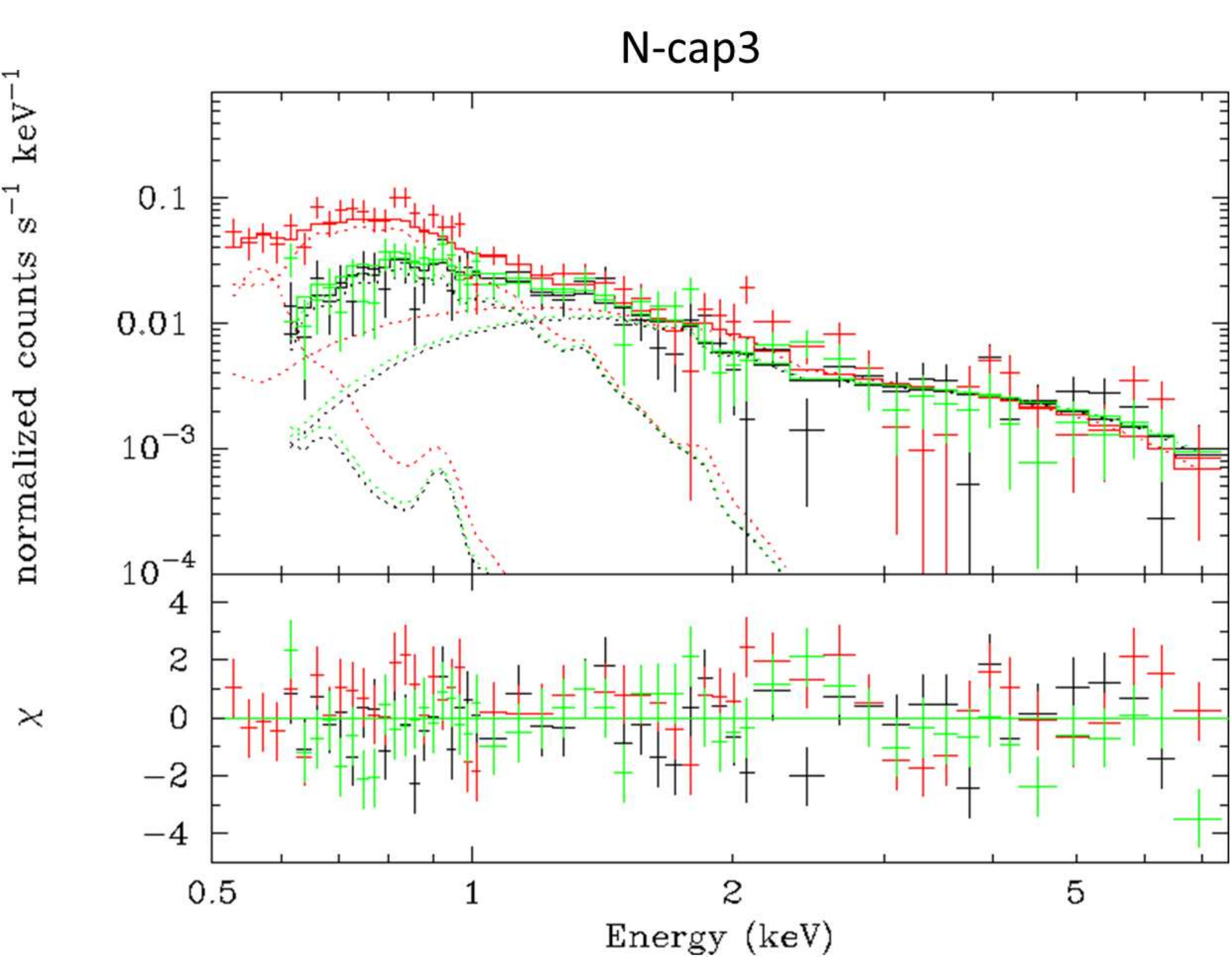}
    \end{minipage}
    \begin{minipage}{0.5\hsize}
      \includegraphics[width=\linewidth]{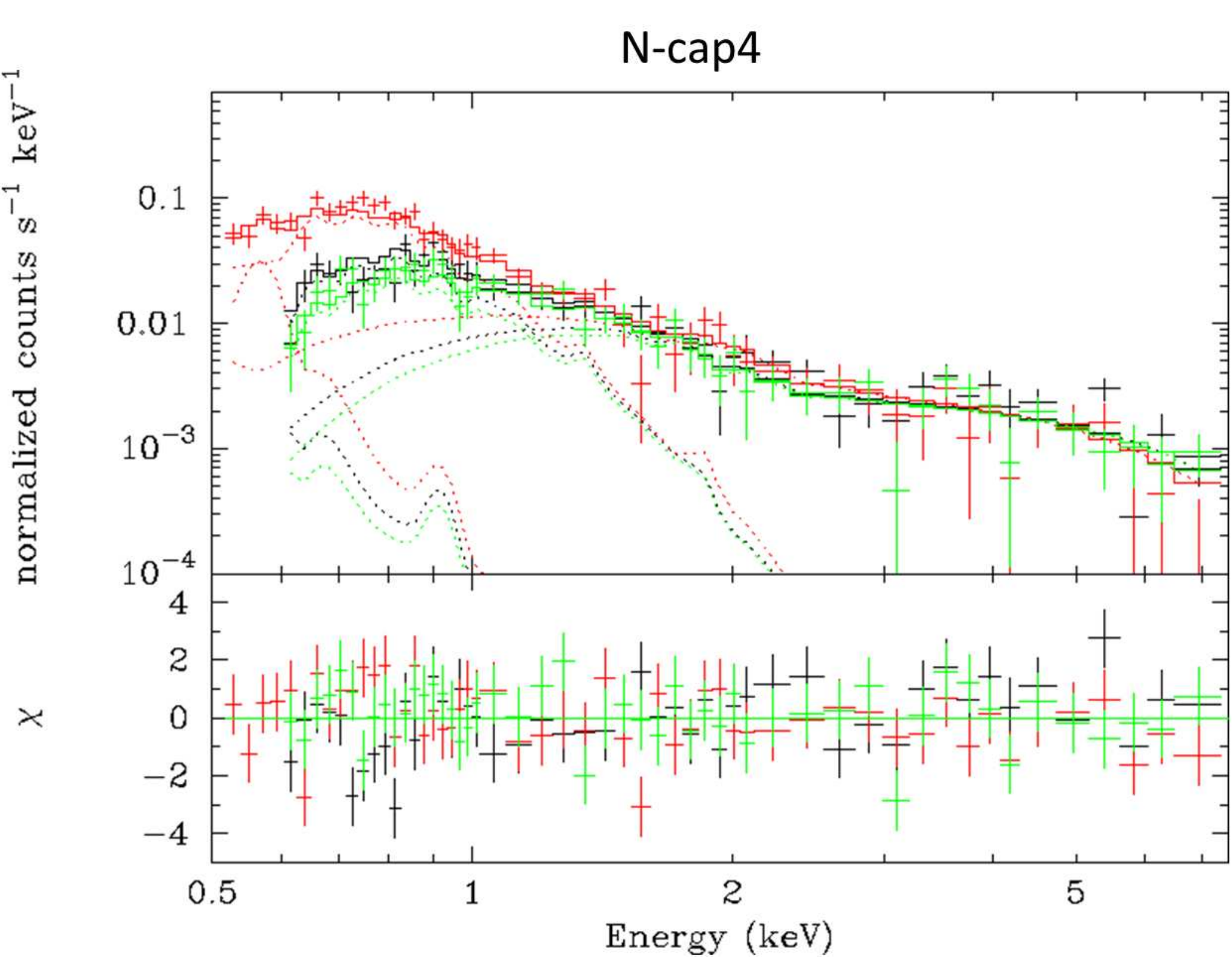}
    \end{minipage}\\
    \begin{minipage}{0.5\hsize}
      \includegraphics[width=\linewidth]{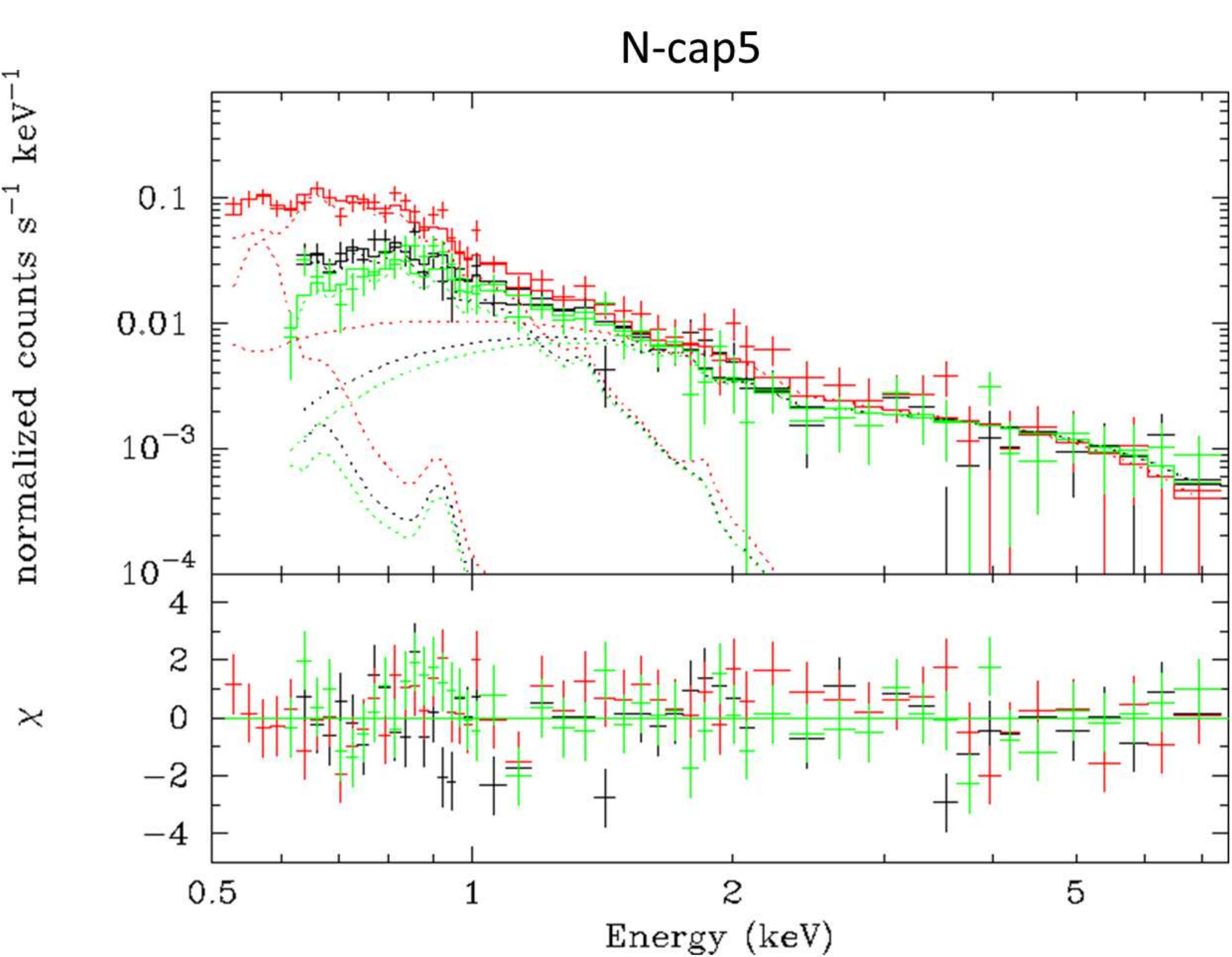}
    \end{minipage}
  \end{tabular}
  \caption{As in figure 3, but for the \textit{Suzaku} archival pointings in the northernmost structure.
    {\it Top-left}: N-cap1.
    {\it Top-right}: N-cap2.
    {\it Middle-left}: N-cap3.
    {\it Middle-right}: N-cap4.
    {\it Bottom}: N-cap5.\label{fig5}}
\end{figure}

\clearpage

\begin{figure}
  \includegraphics[width=\linewidth]{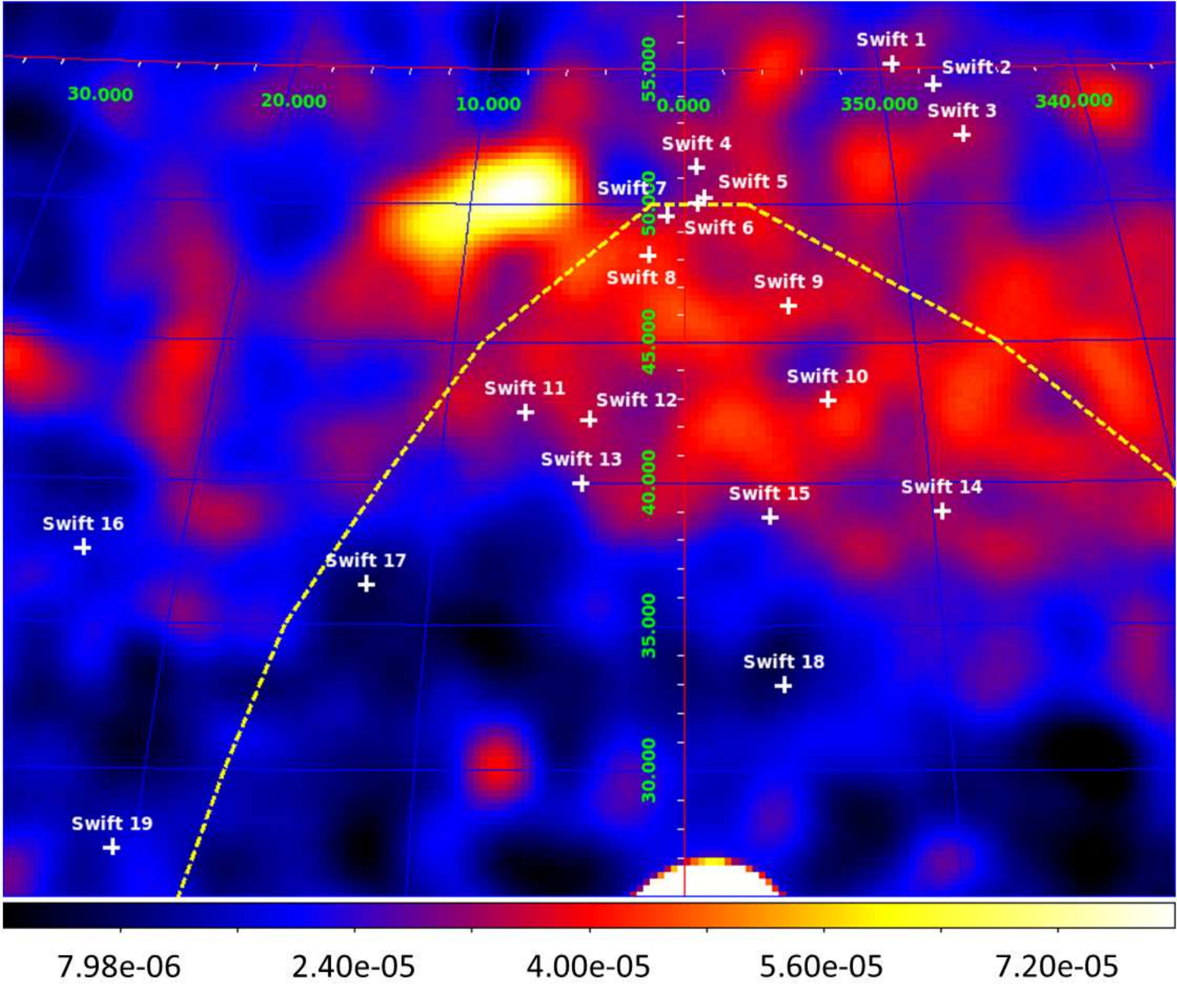}
  \caption{\textit{Swift}-XRT field of views (Swift1-19; white crosses) overlaid with the MAXI-SSC mid-band (1.7$-$4.0 keV) image of the close-up of the northernmost structure. The figure is shown in Galactic coordinates. Yellow dashed lines indicate the boundary of the bubbles as determined by Su et al. (2010). The scale range (units of cts s$^{-1}$ cm$^{-2}$) is indicated at the bottom of the panel.\label{fig6}}
\end{figure}

\clearpage

\begin{figure}
  \includegraphics[width=\linewidth]{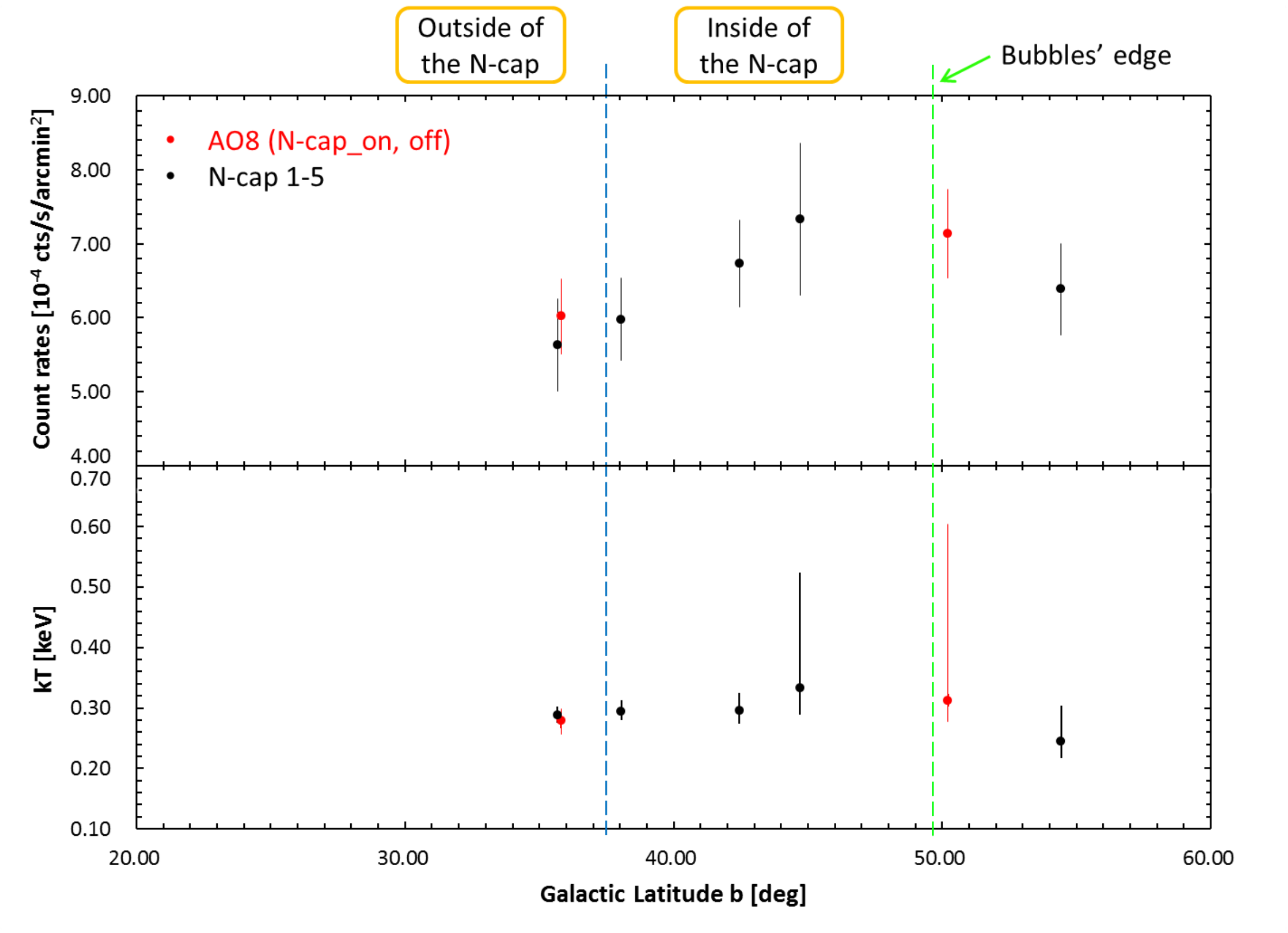}
  \caption{Count rates (1.7$-$4.0 keV) and temperature {\it kT} of the diffuse X-ray emission component APEC2 in the N-cap for our AO8 pointings (N-cap\_on/off; red circles) and the past \textit{Suzaku} observation pointings (N-cap1-5; black circles) plotted against Galactic latitude $b$.\label{fig7}}
\end{figure}

\clearpage

\begin{figure}
  \includegraphics[width=\linewidth]{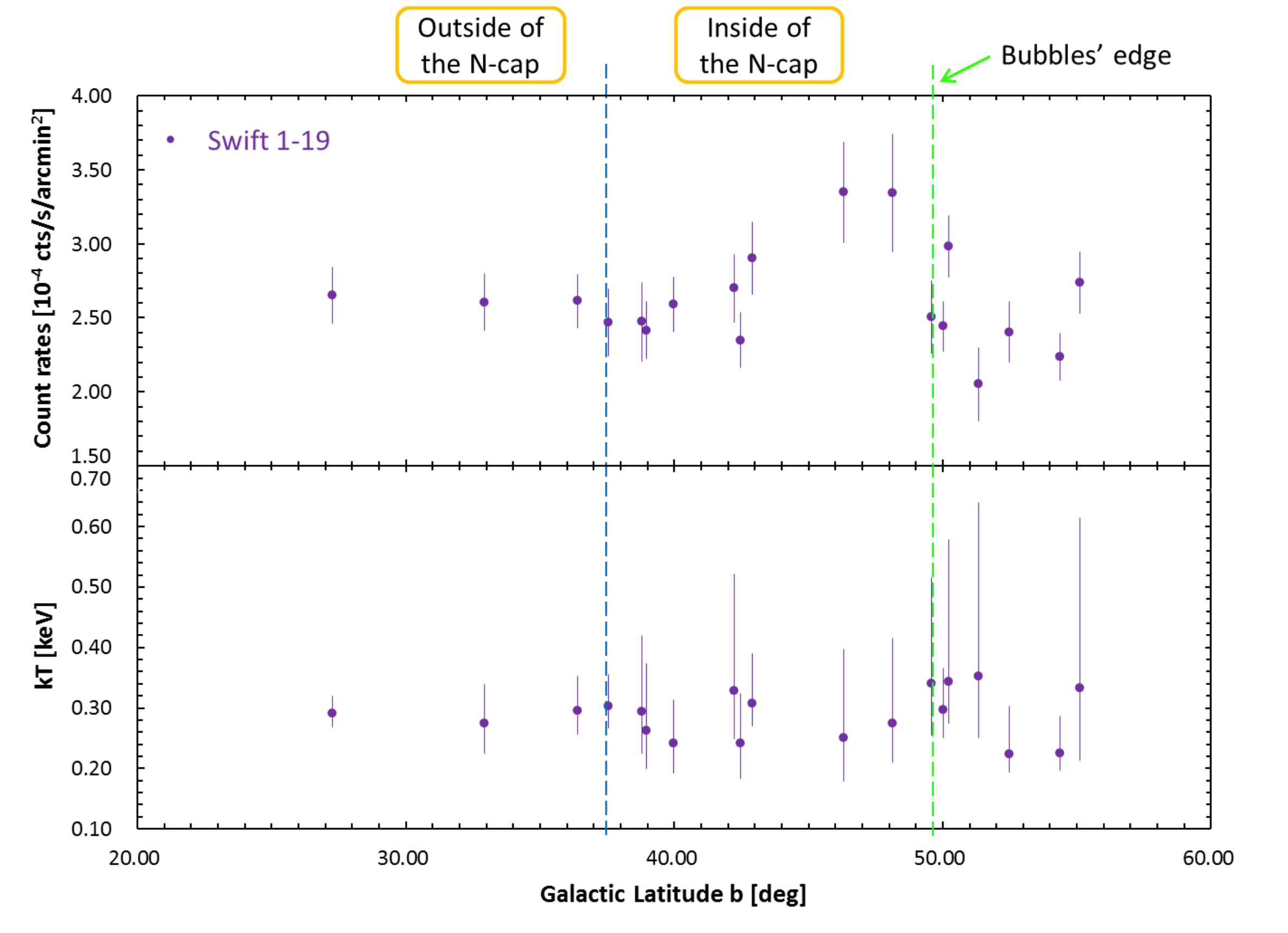}
  \caption{As in figure 7, but for the \textit{Swift} archival pointings (Swift1-19).\label{fig8}}
\end{figure}

\clearpage

\begin{figure}
  \includegraphics[width=\linewidth]{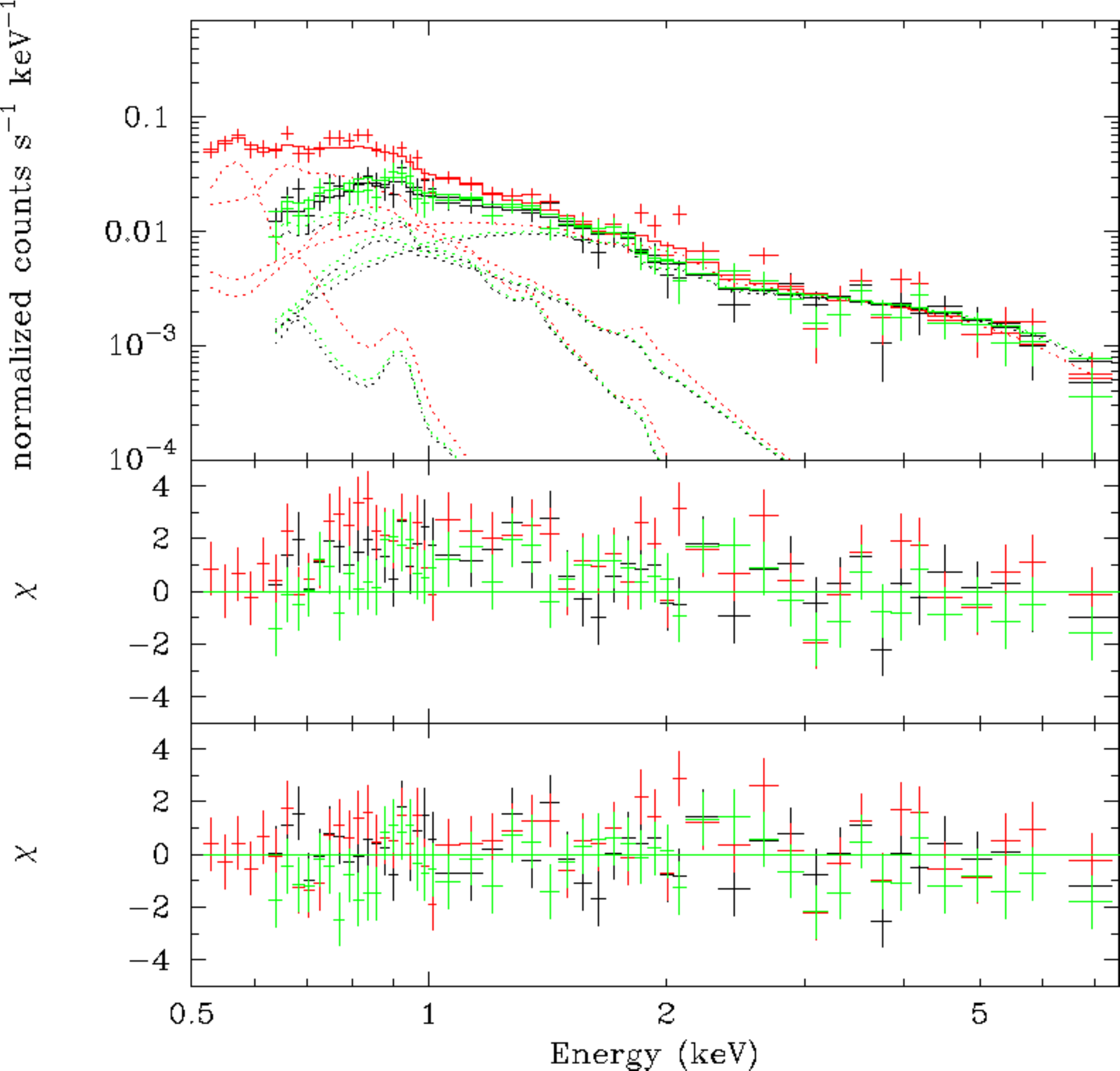}
  \caption{The XIS spectrum of the diffuse emission component for a stack of the data for N-cap\_on, 2 and 3, together with the best-fit model curves 
(APEC1 + WABS *(APEC2 + APEC3 + PL)). The residuals from the model curves without and with the APEC3 component are shown in {\it Middle} and {\it Bottom}, respectively. The LHB component ({\it kT} \(\simeq\) 0.1 keV; APEC1) dominates the lowest 0.5$-$0.6 keV range, the 
GH component ({\it kT} \(\simeq\) 0.3 keV; APEC2) dominates the 0.6$-$1.5 keV range, and a power-law component from the CXB dominates above 1.5 keV 
(\(\Gamma \simeq\) 1.4; PL). In addition, another thermal emission ({\it kT} \(\simeq\) 0.7 keV; APEC3) accounts for the excess of the 1.7$-$4.0 keV 
count rates in the N-cap. XIS 0 data/fits are shown in black, XIS 1 in red, and XIS 3 in green.\label{fig9}}
\end{figure}

\clearpage

\begin{table}[m]
  \small
  \caption{\textit{Suzaku} Observations Log}
  \label{tab:obs_log}

  \begin{center}
    \scalebox{0.9}{
    \begin{tabular}{ccccccccc}
      \hline\hline
      Name & Obs ID & Start time & Stop time & R.A. & Decl. & {\it l} & {\it b} & Exposure \\ 
      & & (UT) & (UT) & [deg]$^{a}$ & [deg]$^{b}$ & [deg]$^{c}$ & [deg]$^{d}$ & [ksec]$^{e}$ \\
      \hline
      N-cap\_on  & 508007010 & 2013/07/26 08:09 & 2013/07/26 20:11 & 221.750 &  -1.312 & 351.952 &  50.223 & 20.7\\
      N-cap\_off & 508008010 & 2013/07/26 20:16 & 2013/07/27 10:14 & 233.686 &  -9.893 & 355.509 &  35.809 & 19.6\\
      SE\_on     & 508009010 & 2013/04/22 16:51 & 2013/04/23 07:56 & 287.398 & -27.250 &   9.973 & -15.747 & 11.8\\
      SE\_off    & 508010010 & 2013/04/23 07:58 & 2013/04/23 19:59 & 288.748 & -25.775 &  11.875 & -16.290 & 16.0\\
      N-cap1     & 807062010 & 2012/08/01 23:39 & 2012/08/02 10:54 & 217.761 &   0.794 & 349.311 &  54.438 & 15.3\\
      N-cap2     & 807058010 & 2012/07/28 08:10 & 2012/07/28 17:58 & 233.434 &   3.616 &   8.894 &  44.702 & 10.4\\
      N-cap3     & 705026010 & 2011/02/01 18:51 & 2011/02/02 04:25 & 230.255 &  -3.837 & 358.141 &  42.451 & 17.5\\
      N-cap4     & 701079010 & 2006/07/19 17:39 & 2006/07/20 15:02 & 220.569 & -17.330 & 337.266 &  38.061 & 32.0\\
      N-cap5     & 401001040 & 2006/02/27 20:38 & 2006/02/28 23:00 & 226.648 & -16.180 & 344.020 &  35.677 & 28.7\\
     \hline
    \end{tabular}
    }
  \end{center}
  $^{a}$: Right ascension of the \textit{Suzaku} pointing center in J2000 equinox.\\
  $^{b}$: Declination of the \textit{Suzaku} pointing center in J2000 equinox.\\
  $^{c}$: Galactic longitude of the \textit{Suzaku} pointing center.\\
  $^{d}$: Galactic latitude of the \textit{Suzaku} pointing center.\\
  $^{e}$: \textit{Suzaku} XIS exposure in ksec.\\
\end{table}

\begin{table}[m]
  \small
  \caption{Fitted parameters of the diffuse emission observed with \textit{Suzaku}}
  \label{tab:fit_param}
  \begin{center}
    \scalebox{0.9}{
    \begin{tabular}{ccccccccc}
      \hline\hline
      Name & $N^{a}_{H,Gal}$ & $N_{H}$/$N^{b}_{H,Gal}$ & {\it kT}$^{c}_{1}$ & EM$^{d}_{1}$ & {\it kT}$^{e}_{2}$ & EM$^{f}_{2}$ & CXB & $\chi^{2}$/dof \\ 
       & (10$^{20}$cm$^{-2}$) &   & (keV) & (10$^{-2}$cm$^{-6}$pc) & (keV) & (10$^{-2}$cm$^{-6}$pc) & Norm$^{g}$ &  \\
      \hline
      N-cap\_on  &  4.12 & \(<7.60\) & 0.1(fix) & 3.54$^{+1.20}_{-1.00}$ & 0.31$^{+0.29}_{-0.04}$ &  5.05$^{+1.47}_{-2.96}$ & 0.93$\pm$0.07 & 179.05/149 \\
      N-cap\_off & 10.69 & \(<0.80\) & 0.1(fix) & 3.32$^{+0.87}_{-0.90}$ & 0.28$\pm$0.02       &  12.2$^{+2.58}_{-1.86}$ & 0.80$\pm$0.06 & 148.01/148 \\
      SE\_on     & 11.87 &  0.5-1.1  & 0.1(fix) & 11.1$\pm$1.84       & 0.31$\pm$0.01       &  8.27$\pm$0.56       & 0.71$\pm$0.08 & 162.56/150 \\
      SE\_off    & 11.56 &  1.7-2.7  & 0.1(fix) & 7.49$\pm$1.18       & 0.31$\pm$0.01       &  3.44$\pm$0.33       & 0.86$\pm$0.06 & 184.38/150 \\
      N-cap1     &  3.02 & \(<12.1\) & 0.1(fix) & 1.66$^{+1.44}_{-1.41}$ & 0.24$^{+0.06}_{-0.03}$ &  6.74$^{+2.94}_{-2.72}$ & 0.79$\pm$0.07 & 186.79/150 \\
      N-cap2     &  4.27 & \(<8.52\) & 0.1(fix) & 5.53$^{+2.16}_{-2.11}$ & 0.33$^{+0.19}_{-0.04}$ &  8.30$^{+2.84}_{-3.90}$ & 0.96$\pm$0.13 & 151.15/150 \\
      N-cap3     &  7.47 &  0.7-4.1  & 0.1(fix) & 1.91$\pm$0.99       & 0.30$^{+0.03}_{-0.02}$ &  9.87$\pm$1.85       & 0.88$\pm$0.07 & 192.43/150 \\
      N-cap4     &  7.82 & \(<2.32\) & 0.1(fix) & 1.17$\pm$0.47       & 0.29$^{+0.02}_{-0.01}$ &  9.51$\pm$1.11       & 0.77$\pm$0.05 & 163.27/150 \\
      N-cap5     &  8.11 & \(<1.82\) & 0.1(fix) & 1.62$\pm$0.52       & 0.28$^{+0.01}_{-0.02}$ &  13.7$^{+1.88}_{-1.33}$ & 0.73$\pm$0.07 & 162.38/149 \\
      \hline 
\end{tabular}
    }
  \end{center}
  {\scriptsize 
    $^{a}$: The absorption column densities for the CXB and the GH components (WABS *(APEC2 + PL)) were fixed to Galactic values given in Dickey \& Lockman (1990).\\
    $^{b}$: The ratio of the absorbing column density to the full Galactic column along the line of sight when $N_{H}$ was left free in the spectral fitting.\\
    $^{c}$: Temperature of the LHB/SWCX plasma fitted with the APEC model for the fixed abundance {\it Z} = {\it Z}$_{\odot}$ .\\
    $^{d}$: Emission measure of the LHB/SWCX plasma fitted with the APEC model for the fixed abundance {\it Z} = {\it Z}$_{\odot}$ .\\
    $^{e}$: Temperature of the GH plasma fitted with the APEC model for the fixed abundance {\it Z} = 0.075 {\it Z}$_{\odot}$ and 0.717 {\it Z}$_{\odot}$ in N-cap and SE-claw, respectively.\\
    $^{f}$: Emission measure of the GH plasma fitted with the APEC model for the fixed abundance {\it Z} = 0.075 {\it Z}$_{\odot}$ and 0.717 {\it Z}$_{\odot}$ in N-cap and SE-claw, respectively.\\
    $^{g}$: The normalization of the CXB in units of 5.85 $\times$ 10$^{-8}$ erg cm$^{-2}$ s$^{-1}$ sr$^{-1}$ (Kushino et al. 2002); see text.\\
  }
\end{table}

\begin{table}[m]
  \small
  \caption{1.7-4.0 keV count rates of the diffuse emission observed with \textit{Suzaku}}
  \label{tab:count-rates}
  \begin{center}
    \scalebox{0.8}{
    \begin{tabular}{ccccccc}
      \hline\hline
      Name & {\it l} $^{a}$ & {\it b} $^{b}$ & count rates (XIS0)$^{c}$ & count rates (XIS1)$^{d}$ & count rates (XIS3)$^{e}$ & count rates (XIS0+1+3)$^{f}$ \\ 
      & [deg] & [deg] & (10$^{-4}$cts s$^{-1}$arcmin$^{-2}$) & (10$^{-4}$cts s$^{-1}$arcmin$^{-2}$) & (10$^{-4}$cts s$^{-1}$arcmin$^{-2}$) & (10$^{-4}$cts s$^{-1}$arcmin$^{-2}$)\\
      \hline
      N-cap\_on  & 351.952 & 50.223 & 2.24$\pm$0.19 & 2.95$\pm$0.22 & 1.95$\pm$0.18 & 7.14$\pm$0.59\\
      N-cap\_off & 355.509 & 35.809 & 1.87$\pm$0.16 & 2.34$\pm$0.18 & 1.81$\pm$0.16 & 6.02$\pm$0.51\\
      N-cap1     & 349.311 & 54.438 & 2.06$\pm$0.20 & 2.28$\pm$0.22 & 2.05$\pm$0.20 & 6.39$\pm$0.62\\
      N-cap2     &   8.894 & 44.702 & 2.14$\pm$0.32 & 3.17$\pm$0.40 & 2.02$\pm$0.31 & 7.33$\pm$0.10\\
      N-cap3     & 358.141 & 42.451 & 1.92$\pm$0.18 & 2.61$\pm$0.21 & 2.21$\pm$0.19 & 6.73$\pm$0.58\\
      N-cap4     & 337.266 & 38.061 & 2.07$\pm$0.17 & 2.04$\pm$0.22 & 1.88$\pm$0.17 & 5.98$\pm$0.56\\
      N-cap5     & 344.020 & 35.677 & 1.71$\pm$0.19 & 2.46$\pm$0.25 & 1.46$\pm$0.18 & 5.63$\pm$0.62\\
      \hline 
\end{tabular}
    }
  \end{center}
  $^{a}$: Galactic longitude of the \textit{Suzaku} pointing center.\\
  $^{b}$: Galactic latitude of the \textit{Suzaku} pointing center.\\
  $^{c}$: 1.7$-$4.0 keV count rates for XIS0 normalized by the area of each ROI.\\
  $^{d}$: 1.7$-$4.0 keV count rates for XIS1 normalized by the area of each ROI.\\
  $^{e}$: 1.7$-$4.0 keV count rates for XIS3 normalized by the area of each ROI.\\
  $^{f}$: The summed 1.7-4.0 keV count rates for XIS0, 1, and 3 normalized by the area of each ROI.\\
\end{table}

\begin{table}[m]
  \small
  \caption{Fitted parameters of the diffuse emission for the stacked \textit{Suzaku} data (N-cap\_on, 2 and 3)}
  \label{tab:fit_param}
  \begin{center}
    \scalebox{0.8}{
    \begin{tabular}{cccccccccc}
      \hline\hline
       & $N^{c}_{H,Gal}$ &  {\it kT}$^{d}_{1}$ & EM$^{e}_{1}$ & {\it kT}$^{f}_{2}$ & EM$^{g}_{2}$ & {\it kT}$^{h}_{3}$ & EM$^{i}_{3}$ & CXB & $\chi^{2}$/dof \\ 
       & (10$^{20}$cm$^{-2}$) & (keV) & (10$^{-2}$cm$^{-6}$pc) & (keV) & (10$^{-2}$cm$^{-6}$pc) & (keV) & (10$^{-2}$cm$^{-6}$pc) & Norm$^{j}$ &  \\
      \hline
      Model 1 $^{a}$ & 5.29 & 0.1(fix) & 3.41$\pm$0.78 & 0.30$\pm$0.02 & 8.64$\pm$1.23 & - & - & 1.04$\pm$0.05 & 168.36/145 \\
      Model 2 $^{b}$ & 5.29 & 0.1(fix) & 3.21$^{+0.91}_{-1.18}$ & 0.26$^{+0.03}_{-0.05}$ & 8.02$^{+2.68}_{-1.88}$ & 0.70$^{+0.22}_{-0.11}$ & 1.13$^{+0.96}_{-0.66}$ & 0.97$\pm$0.06 & 155.73/143 \\
      \hline 
\end{tabular}
    }
  \end{center}
  {\scriptsize 
    $^{a}$: The two APEC model represented by APEC1 + (WABS *(APEC2 + PL)).\\
    $^{a}$: The three APEC model represented by APEC1 + (WABS *(APEC2 + APEC3 + PL)).\\
    $^{c}$: The absorption column densities were fixed to the Galactic values given in Dickey \& Lockman (1990).\\
    $^{d}$: Temperature of the LHB/SWCX plasma fitted with the APEC model for the fixed abundance {\it Z} = {\it Z}$_{\odot}$.\\
    $^{e}$: Emission measure of the LHB/SWCX plasma fitted with the APEC model for the fixed abundance {\it Z} = {\it Z}$_{\odot}$.\\
    $^{f}$: Temperature of the GH plasma fitted with the APEC model for the fixed abundance {\it Z} = 0.075 {\it Z}$_{\odot}$.\\
    $^{g}$: Emission measure of the GH plasma fitted with the APEC model for the fixed abundance {\it Z} = 0.075 {\it Z}$_{\odot}$.\\
    $^{h}$: Temperature of the additional thermal plasma fitted with the APEC model for the fixed abundance {\it Z} = 0.075 {\it Z}$_{\odot}$.\\
    $^{i}$: Emission measure of the additional thermal plasma fitted with the APEC model for the fixed abundance {\it Z} = 0.075 {\it Z}$_{\odot}$.\\
    $^{j}$: The normalization of the CXB in units of 5.85 $\times$ 10$^{-8}$ erg cm$^{-2}$ s$^{-1}$ sr$^{-1}$ (Kushino et al. 2002); see text.\\
  }
\end{table}

\begin{table}[m]
  \small
  \caption{\textit{Swift}-XRT Observations Log}
  \label{tab:obs_log}

  \begin{center}
    \scalebox{0.9}{
    \begin{tabular}{ccccccccc}
      \hline\hline
      Name & Obs ID & Start time & Stop time & R.A. & Decl. & {\it l} & {\it b} & Exposure \\ 
      & & (UT) & (UT) & [deg]$^{a}$ & [deg]$^{b}$ & [deg]$^{c}$ & [deg]$^{d}$ & [ksec]$^{e}$ \\
      \hline
      Swift1  & 00037600001 & 2009/12/19 01:45 & 2009/12/19 08:34 & 217.296 &   1.301 & 349.261 & 55.125 & 5.417\\
      Swift2  & 00037755001 & 2009/12/18 03:11 & 2009/12/18 16:14 & 216.922 &   0.005 & 347.337 & 54.359 & 7.592\\
      Swift3  & 00091308008 & 2013/03/26 06:42 & 2013/03/26 11:56 & 217.608 &  -1.829 & 346.343 & 52.493 & 4.568\\
      Swift4  & 00082093002 & 2013/09/12 14:18 & 2013/09/12 22:12 & 224.263 &   2.802 & 359.375 & 51.352 & 2.615\\
      Swift5  & 00032864005 & 2013/06/30 05:36 & 2013/06/30 18:46 & 224.975 &   1.894 & 359.016 & 50.216 & 5.489\\
      Swift6  & 00033207001 & 2014/03/28 00:14 & 2014/03/28 17:27 & 225.232 &   1.908 & 359.351 & 50.035 & 8.718\\
      Swift7  & 00033265002 & 2014/04/28 10:56 & 2014/04/28 12:39 & 226.089 &   2.310 &   0.777 & 49.575 & 2.979\\
      Swift8  & 00090306002 & 2010/12/25 16:19 & 2010/12/25 18:07 & 227.614 &   1.755 &   1.577 & 48.145 & 1.556\\
      Swift9  & 00090281001 & 2010/04/08 10:32 & 2010/04/08 13:55 & 226.089 &  -2.597 & 355.346 & 46.306 & 2.170\\
      Swift10 & 00036338003 & 2008/01/07 13:54 & 2008/01/07 23:44 & 227.732 &  -5.725 & 353.912 & 42.925 & 5.012\\
      Swift11 & 00039721001 & 2010/12/29 00:45 & 2010/12/30 17:04 & 234.249 &   0.959 &   6.602 & 42.472 & 5.139\\
      Swift12 & 00037942001 & 2008/06/22 07:20 & 2008/06/22 14:02 & 233.223 &  -0.749 &   3.927 & 42.230 & 5.214\\
      Swift13 & 00039723001 & 2011/01/02 01:06 & 2011/01/02 06:16 & 235.090 &  -2.041 &   4.141 & 39.966 & 5.262\\
      Swift14 & 00040980001 & 2010/09/16 01:01 & 2010/09/16 07:44 & 228.007 & -10.863 & 349.572 & 38.957 & 5.439\\
      Swift15 & 00055750014 & 2011/01/14 03:35 & 2011/01/14 07:05 & 232.086 &  -7.240 & 356.528 & 38.790 & 2.555\\
      Swift16 & 00035800002 & 2006/10/08 16:10 & 2006/10/09 05:09 & 245.412 &   9.558 &  23.816 & 37.542 & 4.580\\
      Swift17 & 00037281001 & 2008/01/20 00:44 & 2008/01/20 16:59 & 241.792 &   1.112 &  12.405 & 36.399 & 8.693\\
      Swift18 & 00036065002 & 2007/01/17 00:55 & 2007/01/17 23:33 & 236.199 & -11.491 & 356.179 & 32.920 & 6.158\\
      Swift19 & 00037283001 & 2008/01/20 18:24 & 2008/01/21 13:46 & 253.272 &   2.403 &  20.746 & 27.269 & 8.029\\
     \hline
    \end{tabular}
    }
  \end{center}
  $^{a}$: Right ascension of the \textit{Swift} pointing center in J2000 equinox.\\
  $^{b}$: Declination of the \textit{Swift} pointing center in J2000 equinox.\\
  $^{c}$: Galactic longitude of the \textit{Swift} pointing center.\\
  $^{d}$: Galactic latitude of the \textit{Swift} pointing center.\\
  $^{e}$: \textit{Swift}-XRT exposure in ksec.\\
\end{table}

\begin{table}[m]
  \small
  \caption{Fitted parameters of the diffuse emission observed with \textit{Swift}-XRT}
  \label{tab:fit_param}
  \begin{center}
    \scalebox{0.9}{
    \begin{tabular}{cccccccc}
      \hline\hline
      Name & $N^{a}_{H,Gal}$      & {\it kT}$^{b}_{1}$ & EM$^{c}_{1}$          & {\it kT}$^{d}_{2}$ & EM$^{e}_{2}$           & CXB       & $\chi^{2}$/dof \\ 
         & (10$^{20}$cm$^{-2}$) & (keV)            & (10$^{-2}$cm$^{-6}$pc) & (keV)             & (10$^{-2}$cm$^{-6}$pc) & Norm$^{f}$ &  \\
      \hline
      Swift1  &  2.85 & 0.1(fix) & 0.59$^{+0.23}_{-0.35}$ & 0.33$^{+0.28}_{-0.12}$ & 0.27$^{+0.27}_{-0.17}$ & 2.00$\pm$0.21 & 1.08/39 \\
      Swift2  &  3.09 & 0.1(fix) & \(<0.37\)           & 0.23$^{+0.06}_{-0.03}$ & 0.77$^{+0.21}_{-0.37}$ & 1.56$\pm$0.15 & 1.58/39 \\
      Swift3  &  3.75 & 0.1(fix) & \(<0.47\)           & 0.22$^{+0.08}_{-0.03}$ & 1.04$^{+0.27}_{-0.54}$ & 1.81$\pm$0.20 & 1.23/39 \\
      Swift4  &  3.91 & 0.1(fix) & 1.77$^{+0.48}_{-0.49}$ & 0.35$^{+0.29}_{-0.10}$ & 0.49$^{+0.39}_{-0.30}$ & 1.64$\pm$0.27 & 1.33/39 \\
      Swift5  &  4.19 & 0.1(fix) & 1.10$^{+0.33}_{-0.38}$ & 0.34$^{+0.24}_{-0.07}$ & 0.73$^{+0.35}_{-0.34}$ & 2.31$\pm$0.22 & 0.77/39 \\
      Swift6  &  4.22 & 0.1(fix) & 0.38$^{+0.26}_{-0.30}$ & 0.30$^{+0.07}_{-0.05}$ & 0.96$^{+0.35}_{-0.28}$ & 1.63$\pm$0.21 & 1.16/39 \\
      Swift7  &  4.15 & 0.1(fix) & 2.10$^{+0.44}_{-0.64}$ & 0.34$^{+0.18}_{-0.09}$ & 0.83$^{+0.57}_{-0.35}$ & 1.92$\pm$0.27 & 1.03/39 \\
      Swift8  &  3.99 & 0.1(fix) & 1.03$^{+0.73}_{-0.84}$ & 0.28$^{+0.14}_{-0.06}$ & 1.18$^{+0.92}_{-0.64}$ & 1.99$\pm$0.38 & 1.24/38 \\
      Swift9  &  5.98 & 0.1(fix) & 1.26$^{+0.66}_{-1.20}$ & 0.25$^{+0.15}_{-0.07}$ & 1.14$^{+2.04}_{-0.71}$ & 2.54$\pm$0.35 & 0.88/39 \\
      Swift10 &  6.80 & 0.1(fix) & \(<0.23\)           & 0.31$^{+0.08}_{-0.04}$ & 1.05$^{+0.21}_{-0.33}$ & 2.24$\pm$0.26 & 0.75/39 \\
      Swift11 &  5.21 & 0.1(fix) & 1.19$^{+0.50}_{-0.88}$ & 0.24$^{+0.08}_{-0.06}$ & 1.16$^{+1.48}_{-0.56}$ & 1.81$\pm$0.19 & 0.80/39 \\
      Swift12 &  6.35 & 0.1(fix) & 0.37$^{+0.21}_{-0.28}$ & 0.33$^{+0.19}_{-0.08}$ & 0.51$^{+0.33}_{-0.24}$ & 1.95$\pm$0.25 & 1.29/39 \\
      Swift13 &  7.78 & 0.1(fix) & 0.65$^{+0.46}_{-0.65}$ & 0.24$^{+0.07}_{-0.05}$ & 1.42$^{+1.30}_{-0.64}$ & 2.34$\pm$0.21 & 1.06/39 \\
      Swift14 &  8.84 & 0.1(fix) & 0.86$^{+0.37}_{-0.51}$ & 0.26$^{+0.11}_{-0.06}$ & 1.03$^{+1.04}_{-0.50}$ & 2.00$\pm$0.21 & 0.84/39 \\
      Swift15 &  7.74 & 0.1(fix) & 0.72$^{+0.44}_{-0.52}$ & 0.29$^{+0.13}_{-0.07}$ & 1.02$^{+0.74}_{-0.48}$ & 1.82$\pm$0.30 & 0.85/37 \\
      Swift16 &  4.50 & 0.1(fix) & 1.44$\pm$0.52       & 0.30$^{+0.05}_{-0.04}$ & 1.94$^{+0.52}_{-0.51}$ & 1.82$\pm$0.24 & 1.13/39 \\
      Swift17 &  6.66 & 0.1(fix) & 0.75$^{+0.27}_{-0.28}$ & 0.30$^{+0.06}_{-0.04}$ & 1.09$^{+0.36}_{-0.31}$ & 1.90$\pm$0.19 & 1.19/39 \\
      Swift18 & 12.23 & 0.1(fix) & 0.48$^{+0.25}_{-0.32}$ & 0.27$^{+0.07}_{-0.05}$ & 1.29$^{+0.73}_{-0.44}$ & 1.99$\pm$0.21 & 1.19/39 \\
      Swift19 &  5.70 & 0.1(fix) & 0.92$\pm$0.39       & 0.29$^{+0.03}_{-0.02}$ & 2.72$^{+0.45}_{-0.45}$ & 2.03$\pm$0.22 & 1.30/39 \\
      \hline 
\end{tabular}
    }
  \end{center}
  {\scriptsize 
    $^{a}$: The absorption column densities for the CXB and the GH components (WABS *(APEC2 + PL)) were fixed to Galactic values given in Dickey \& Lockman (1990).\\
    $^{b}$: Temperature of the LHB/SWCX plasma fitted with the APEC model for the fixed abundance {\it Z} = {\it Z}$_{\odot}$.\\
    $^{c}$: Emission measure of the LHB/SWCX plasma fitted with the APEC model for the fixed abundance {\it Z} = {\it Z}$_{\odot}$.\\
    $^{d}$: Temperature of the GH plasma fitted with the APEC model for the fixed abundance {\it Z} = 0.2 {\it Z}$_{\odot}$.\\
    $^{e}$: Emission measure of the GH plasma fitted with the APEC model for the fixed abundance {\it Z} = 0.2 {\it Z}$_{\odot}$.\\
    $^{f}$: The normalization of the CXB in units of 5.85 $\times$ 10$^{-8}$ erg cm$^{-2}$ s$^{-1}$ sr$^{-1}$ (Kushino et al. 2002); see text.\\
  }
\end{table}

\begin{table}[m]
  \small
  \caption{1.7$-$4.0 keV count rates of the diffuse emission observed with \textit{Swift}-XRT}
  \label{tab:count-rates}
  \begin{center}
    {
    \begin{tabular}{ccccc}
      \hline\hline
      Name & {\it l} $^{a}$ & {\it b} $^{b}$ & XRT count rates$^{c}$            & window size$^{d}$\\ 
         & [deg]         & [deg]          & (10$^{-4}$cts s$^{-1}$arcmin$^{-2}$) & (pixel)\\
      \hline
      Swift1  & 349.261 & 55.125 & 2.74$\pm$0.21 & 600 x 600\\
      Swift2  & 347.337 & 54.359 & 2.24$\pm$0.16 & 600 x 600\\
      Swift3  & 346.343 & 52.493 & 2.40$\pm$0.20 & 600 x 600\\
      Swift4  & 359.375 & 51.352 & 2.05$\pm$0.25 & 600 x 600\\
      Swift5  & 359.016 & 50.216 & 2.98$\pm$0.21 & 600 x 600\\
      Swift6  & 359.351 & 50.035 & 2.44$\pm$0.17 & 600 x 600\\
      Swift7  &   0.777 & 49.575 & 2.51$\pm$0.25 & 600 x 600\\
      Swift8  &   1.577 & 48.145 & 3.34$\pm$0.40 & 600 x 600\\
      Swift9  & 355.346 & 46.306 & 3.35$\pm$0.34 & 600 x 600\\
      Swift10 & 353.912 & 42.925 & 2.90$\pm$0.24 & 500 x 500\\
      Swift11 &   6.602 & 42.472 & 2.35$\pm$0.19 & 600 x 600\\
      Swift12 &   3.927 & 42.230 & 2.70$\pm$0.23 & 600 x 600\\
      Swift13 &   4.141 & 39.966 & 2.59$\pm$0.19 & 600 x 600\\
      Swift14 & 349.572 & 38.957 & 2.42$\pm$0.19 & 600 x 600\\
      Swift15 & 356.528 & 38.790 & 2.47$\pm$0.27 & 600 x 600\\
      Swift16 &  23.816 & 37.542 & 2.47$\pm$0.23 & 500 x 500\\
      Swift17 &  12.405 & 36.399 & 2.61$\pm$0.18 & 500 x 500\\
      Swift18 & 356.179 & 32.920 & 2.61$\pm$0.20 & 500 x 500\\
      Swift19 &  20.746 & 27.269 & 2.65$\pm$0.19 & 500 x 500\\
      \hline 
\end{tabular}
    }
  \end{center}
  $^{a}$: Galactic longitude of the \textit{Swift} pointing center.\\
  $^{b}$: Galactic latitude of the \textit{Swift} pointing center.\\
  $^{c}$: 1.7$-$4.0 keV count rates for XRT normalized by the area of each ROI.\\
  $^{d}$: The window setting of the XRT CCD.\\
\end{table}

\end{document}